\documentclass[11pt]{article}
\usepackage{epic}
\usepackage{curves}
\catcode`\@=11
\def\marginnote#1{}

\newcount\hour
\newcount\minute
\newtoks\amorpm
\hour=\time\divide\hour by60
\minute=\time{\multiply\hour by60 \global\advance\minute by-\hour}
\edef\standardtime{{\ifnum\hour<12 \global\amorpm={am}%
        \else\global\amorpm={pm}\advance\hour by-12 \fi
        \ifnum\hour=0 \hour=12 \fi
        \number\hour:\ifnum\minute<10 0\fi\number\minute\the\amorpm}}
\edef\militarytime{\number\hour:\ifnum\minute<10 0\fi\number\minute}

\def\draftlabel#1{{\@bsphack\if@filesw {\let\thepage\relax
      \xdef\@gtempa{\write\@auxout{\string
          \newlabel{#1}{{\@currentlabel}{\thepage}}}}}\@gtempa \if@nobreak
    \ifvmode\nobreak\fi\fi\fi\@esphack} \gdef\@eqnlabel{#1}}
    \def\@eqnlabel{}
\def\@vacuum{}
\def\draftmarginnote#1{\marginpar{\raggedright\scriptsize\tt#1}}

\def\draft{
  \oddsidemargin -.5truein
  \def\@oddfoot{\footnotesize \sl preliminary draft \hfil
    \rm\thepage\hfil\sl\today\quad\militarytime}
  \let\@evenfoot\@oddfoot \overfullrule 3pt
    \let\label=\draftlabel
    \let\marginnote=\draftmarginnote
  \def\@eqnnum{(\theequation)\rlap{\kern\marginparsep\tt\@eqnlabel}%
    \global\let\@eqnlabel\@vacuum}

  }

\voffset= - 1.2in
\hoffset= - 1.0in

\def\be{\begin{equation}}
\def\ee{\end{equation}}
\def\bea{\begin{eqnarray}}
\def\eea{\end{eqnarray}}
\def\be{\ba}
\def\ee{\ea}
\def\<{\langle}
\def\>{\rangle}
\def\stackreb#1#2{\mathrel{\mathop{#2}\limits_{#1}}}

\def\res{{{\rm res}}}

\def\F{{\cal F}}

\def\d{\partial}
\def\N2{${\cal N}=2$}

\def\tr{{\mathrm{tr\,}}}
\def\diag{{\rm diag}}
\def\1N{${\cal N}=1$}
\def\4N{${\cal N}=4$}

\def\e{{\,\rm e}\,}

\def\bea{\begin{eqnarray}}
\def\eea{\end{eqnarray}}
\def\bqa{\begin{eqnarray}}
\def\eqa{\end{eqnarray}}

\def\beq{\begin{equation}}
\def\eeq{\end{equation}}
\def\ba{\beq\begin{array}{c}}
\def\ea{\end{array}\eeq}
\def\be{\beq}
\def\ee{\eeq}

\let\operatorname=\mathrm
\let\text=\mathrm

\newcommand\theTag[1]{(\ref{#1})}

\def\ch{\operatorname{cosh}}

\def\diag{\operatorname{diag}}
\def\e{e}

\def\beq{\begin{equation}}
\def\eeq{\end{equation}}
\def\bea{\begin{eqnarray}}
\def\eea{\end{eqnarray}}

\newcommand{\rf}[1]{(\ref{#1})}

\def\t{\widetilde}

\def\F {{\cal F}}

\def\m {\mu}

\renewcommand{\d}{{{\partial}}}

\renewcommand{\<}{\langle}
\renewcommand{\>}{\rangle}

\def\diag{\hbox{diag\,}}

\def\2{{1\over 2}}
\def\stackreb#1#2{\mathrel{\mathop{#2}\limits_{#1}}}

\def\d{\partial}

\def\â{$\tau$}

\newcommand{\lm}{\lambda}

\newcommand{\cpict}[3]{
\dimen1=#1\advance\dimen1 by-\hsize\divide\dimen1 by-2
\vtop to #2{
\noindent\hskip\dimen1{\special{em:graph #3.bmp}}
\vfil}\hskip-2cm
}

\newcommand{\dV}{\frac{\partial}{\partial V(\lambda)}}
\newcommand{\dmul}{\frac{\partial\mu_\alpha}{\partial V(\lambda)}}
\newcommand{\parV}{\frac{\partial}{\partial V(\lambda)}}
\newcommand{\cI}{\oint_{\cal C_{\cal D}}\frac{d\lambda}{2\pi i}}
\newcommand{\Vp}{V^{\prime}}

\newcommand{\ty}{{\tilde y}}

\let\@@savethanks\thanks
\def\thanks#1{\gdef\thefootnote{\alph{footnote}}\@@savethanks{#1}}

\title{{\bf Complex Geometry of Matrix Models
}
\vspace{.5cm}}
\author{{\bf L. Chekhov}\thanks{E-mail: \ chekhov@mi.ras.ru}
\date{ } \\ {\small
{\it Steklov Mathematical Institute, ITEP}, and {\it LIFR, Moscow, Russia}}\\ \\
{\bf A. Marshakov}\footnote{E-mail: \ mars@lpi.ru; mars@itep.ru
}\ \ \ {\bf A. Mironov}\footnote{E-mail:
\ mironov@itep.ru; mironov@lpi.ru}
\date{ } \\
{\small {\it Lebedev Physics Institute}
and {\it ITEP, Moscow, Russia}}\\ \\
{\bf D.Vasiliev}\thanks{E-mail: \ vasiliev@itep.ru}
\date{ } \\ {\small {\it MIPT, Moscow, Russia} and
{\it ITEP, Moscow, Russia}}
}

\begin{document}

\maketitle

\vspace{-10.5cm}

\begin{center}
\hfill FIAN/TD-16/04\\
\hfill ITEP/TH-48/04\\
\end{center}

\vspace{7.5cm}

\begin{abstract}
\noindent
The paper contains some new results and a review of recent
achievements, concerning the multisupport solutions to matrix models.
In the leading order of the 't Hooft expansion for matrix integral,
these solutions are described by quasiclassical or generalized Whitham hierarchies
and are directly related to the superpotentials of four-dimensional
${\cal N}=1$ SUSY gauge theories. We study the derivatives of tau-functions
for these solutions, associated with
the families of Riemann surfaces (with possible double points), and relations
for these derivatives imposed by complex geometry, including the
WDVV equations. We also
find the free energy in subleading order of the 't Hooft expansion and
prove that it satisfies certain determinant relations.
\end{abstract}
\def\thefootnote{\arabic{footnote}}

Recent interest to matrix models and especially to their so-called
multisupport (multicut) solutions was inspired by the studies in
${\cal N}=1$
SUSY gauge
theories due to Cachazo, Intrilligator and Vafa \cite{CIV1}, \cite{CIV2}
and by the proposal of Dijkgraaf and Vafa \cite{DV} to calculate the
low energy superpotentials, using the partition function of multicut solutions.
The solutions themselves are well-known already for a long time
(see, e.g., \cite{JU90,AkAm}) with a new vim due to the paper by Bonnet, David and Eynard
\cite{David}.

The Dijkgraaf--Vafa proposal was to consider the nonperturbative
superpotentials of ${\cal N}=1$ SUSY gauge theories in
four dimensions (possibly coming as the softly broken ${\cal N}=2$
Seiberg--Witten (SW) theories \cite{WS1,WS2}) arising from the partition
functions of the one-matrix model (1MM) in the leading order in
$1/N$, $N$ being the matrix size. The leading order (of the 't Hooft
$1/N$-expansion) of the matrix model is described by the
quasiclassical tau-function of the so-called universal Whitham
hierarchy \cite{Kri1} (see also \cite{RG,MarMir98,Mbook,SWbook,GM}, the
details about one-matrix and two-matrix cases see in \cite{ChM} and
\cite{KM}). One would expect the existence of the relation between the partition
function in the planar limit and the quasiclassical, or Whitham hierarchy
already because matrix integrals are tau-functions of the
hierarchies of integrable equations of the KP/Toda type
\cite{IntMamo}. For the planar single-cut solutions the matrix model,
partition functions become tau-functions of the dispersionless
Toda hierarchy, one of the simplest example of the Whitham
hierarchy.

One may also consider more general solutions to matrix models, identifying
them with generic solutions to the loop (Schwinger--Dyson, or Virasoro)
equations \cite{48}, to be the Ward identities satisfied by matrix integrals
\cite{Virasoro}.
These generalized solutions to the loop equations can be still treated as
matrix model partition functions (e.g., within a D-module
ideology \cite{AMM,AMM2,AMM3}); however, they do not necessarily admit
any matrix integral representation.

In what follows, we aim to discuss an interesting class of
multi-cut, or multi-support, solutions to the loop equations that
{\it have} multi-matrix integral representation
\cite{David,KMT,AMM,AMM3}. These solutions are associated with families
of Riemann surfaces and form a sort of a basis in the space of all
solutions to the loop equations \cite{AMM,AMM3} (like the finite-gap
solutions form a basis in the space of all solutions to an
integrable hierarchy). They can be distinguished by their
``isomonodromic" properties---switching on higher matrix model
couplings, or $1/N$-corrections does not change the family of
Riemann surfaces, but just reparameterizes the moduli as functions
of these couplings. This property is directly related to that
the partition functions of these solutions are quasiclassical
tau-functions (also often called as prepotentials of the corresponding
Seiberg--Witten-like systems).

In sect.~\ref{s:1MM}, we describe the general properties of multi-cut solutions of
matrix models. In particular, we prove that the free energy of
the 1MM in the planar (large $N$) limit coincides with the
prepotential of some Seiberg--Witten-like theory. This free
energy is the logarithm of a quasiclassical tau-function.
The corresponding quasiclassical hierarchy is explicitly constructed.

The Whitham hierarchy is basically formulated in terms of Abelian differentials on
a family of Riemann surfaces \cite{Kri1}. This implies the main quantities in matrix
models are to be expressed in geometric terms and allow
calculating derivatives of the matrix model free energy. Indeed, we demonstrate
in sect.~\ref{s:geometry} that the
second derivatives of the logarithm of the matrix model partition function
can be expressed through the so-called Bergmann bi-differential.

In sect.~\ref{ss:WDVV}, we turn to the third derivatives of partition
function and to the
Witten--Dijkgraaf--Verlinde--Verlinde
(WDVV) equations \cite{Wit90,LGMT,wdvvg}, which are differential equations,
involving the third derivatives of tau-function with respect to Whitham times.
These equations are usually considered an evidence for existing an
underlying topological string theory. In sect.~3, we prove
that the quasiclassical tau-function of the multi-support solutions
to matrix models
satisfies the WDVV equations in the case of general 1MM solution, i.e., in
the case of arbitrary number of nonzero times,
which include now, besides the times of the potential, the occupation numbers
(filling fractions)
indicating the portions of eigenvalues of the corresponding model that
dwell on the
related intervals of eigenvalue supports.
Although being, at first glance, quantities
of very different nature in comparison with the original times,
they can be nicely combined into a
unified set of the "small phase space" of the model
\footnote{Note that, from the point of view of ${\cal N}=1$ SUSY theory,
couplings in the matrix model potential must be
identified with couplings in the tree
superpotential, while the occupation numbers play the role of moduli
of the loop equation solutions and must be associated with the vacuum
expectation values of the gluino condensates.}.
This completes an interpretation of the results of
\cite{CIV1,CIV2} in terms of quasiclassical hierarchies.

The WDVV equations are a simple consequence
of the residue formula and associativity of some algebra (e.g., of the
holomorphic differentials on the Riemann surface) \cite{wdvvg,wdvv2,wdvvlong}
(see also \cite{wdvvmore}); moreover, the associativity of algebra can be replaced
by a simple counting argument \cite{MaWDVV}: the number of critical points in the
residue formula should be equal to the dimension of the "small phase space".
We present the proof of the residue formula and extra conditions following
\cite{ChMMV}. Note
that, while the residue formula is always present in the theories of such
type \cite{Kri1,wdvvlong},\footnote{The recently proposed in \cite{Ey}
residue formulas in planar case just follow from
the standard residue formula. However, the residue formulas of \cite{Ey}
involving non-planar corrections look new and rather instructive.}
the associativity (naive counting)
is often violated \cite{wdvvlong} (\cite{MaWDVV})
(maybe the most notorious
case is the SW system associated with the elliptic
Calogero--Moser model, where this phenomenon was first found in \cite{wdvvlong}).

There are strong indications that the correspondence
between matrix models and SUSY gauge theories goes beyond
just the large-$N$ limit of matrix models. Say, the
relation between gauge theories and matrix model were further verified at the
nonplanar, genus-one level for the solution with two cuts and a cubic matrix model
potential~\cite{KMT}.
In sect.~\ref{s:genus}, we
find the multicut solution to 1MM in the subleading order: the torus
approximation in terms of a dual string theory. We calculate the corresponding free
energy and,
in particular, prove it to have a determinant form, related to the topological
B-model on the local geometry $\widehat{II}$
and conjectured in \cite{DV} (the authors of \cite{KMT} have checked it for
the several first terms of expansion in the
two-cut case).
Then, we apply formulas for the third derivatives of the planar
free energy and for the genus-one free energy obtained in the paper,
and conjecture a diagrammatic interpretation of
these formulas to be extended to higher genera.

\section{Matrix models and generalized Whitham hierarchies \label{s:1MM}}

\subsection{Matrix integrals and resolvents \label{ss:resolvent}}

Consider the 1MM integral\footnote{There is a more consistent
point of view on matrix integrals and loop equations below
that implies introducing some background
(polynomial) potential $V_0(x)$ into the exponential (\ref{Xap2.1}), i.e.,
shifting first several $t_k\to T_k+t_k$ and then looking at this matrix
integral as a formal series in $t_k$'s (see, e.g., \cite{GKMU,AMM,AMM3} for a review).
Then, the loop equations are iteratively solved. Such a framework is more
effective when looking at the whole manifold of solutions to the loop
equations. However, this is at the price of rigid fixing the number of
cuts in the multicut solution from the very beginning. Since we will freely
change the number of cuts (which is a smooth procedure specifically in the multicut
solutions) and do not care of other than multicut solutions to the loop
equations, we follow a less pure scheme (see, e.g., \cite{JU90,David,Ak96} and
references therein) that basically would imply some re-summation of
infinite series, etc. It allows us, however, instead of dealing with infinite
series to deal with objects determined on Riemann surfaces.}
\be
\int_{N\times N}DX\, \e^{-{1\over \hbar}\tr V(X)}=\e^{\cal F},
\label{Xap2.1}
\ee
where $V(X)=\sum_{n\geq 1}^{}t_nX^n$, $\hbar = {t_0\over N}$ is
a formal expansion parameter, the integration goes
over the $N\times N$ matrices, $DX\propto\prod_{ij}dX_{ij}$, and
for generic potential one should consider the
{\em holomorphic} version of \rf{Xap2.1} implying contour integration in complex plane
in each variable.
The topological expansion of the Feynman diagrams series is then equivalent to
the expansion in even powers of $\hbar$ for
\be
{\cal F}\equiv {\cal F}(\hbar,t_0, t_1, t_2, \dots)
=\sum_{h=0}^{\infty}{\hbar}^{2h-2}{\cal F}_h,
\label{Xap2.2}
\ee
Customarily $t_0=\hbar N$ is the scaled
number of eigenvalues. We assume the potential $V(p)$ to be a polynomial
of the fixed degree $m+1$, with the fixed constant "highest" time $t_{m+1}$.

The averages, corresponding to the partition function~\theTag{Xap2.1} are
defined as usual:
\beq
\bigl\langle f(X)\bigr\rangle=
\frac1Z\int_{N\times N}DX\,f(X)\,\exp\left(-{1\over \hbar}\tr V(X)\right)
\label{4.1}
\eeq
and it is convenient to use their
generating functionals: the one-point resolvent
\beq
W(\lambda)=
\hbar
\sum_{k=0}^{\infty}
\frac{\langle\tr X^{k}\rangle}{\lambda^{k+1}}
\label{4.2}
\eeq
as well as the $s$-point resolvents $(s\geq2)$
\beq
W(\lambda_1,\dots,\lambda_s)=
\hbar^{2-s}
\sum_{k_1,\dots,k_s=1}^{\infty}
\frac{\langle\tr X^{k_1}\cdots\tr X^{k_s}\rangle_{\mathrm{conn}}}
{\lambda_1^{k_1+1}\cdots \lambda_s^{k_s+1}}=
\hbar^{2-s}
\left\langle\tr\frac{1}{\lambda_1-X}\cdots
\tr\frac{1}{\lambda_s-X}\right\rangle_{\mathrm{conn}}
\label{4.3}
\eeq
where the subscript ``$\mathrm{conn}$" pertains to the connected
part.

These resolvents are obtained from the free energy ${\cal F}$ through the
action
\bea
W(\lambda_1,\dots,\lambda_s)&=&\hbar^2\frac{\d}{\d V(\lambda_s)}\frac{\partial}{\partial V(\lambda_{s-1})}\cdots
\frac{\partial {\cal F}}{\partial V(\lambda_1)}=\nonumber
\\
&=&\frac{\partial }{\partial V(\lambda_s)}\frac{\partial }{\partial V(\lambda_{s-1})}\cdots
\frac{\partial }{\partial V(\lambda_2)}W(\lambda_1),
\label{4.5}
\eea
of the loop insertion operator\footnote{This operator contains all
partial derivatives w.r.t. the variables
$t_k$'s. However, below we introduce additional variables
$S_i$ and, therefore, we use the
partial derivative notation here.}
\beq
\frac{\partial }{\partial V(\lambda)}\equiv
-\sum_{j=1}^{\infty}\frac{1}{\lambda^{j+1}}\frac{\d}{\d t_{j}}.
\label{4.6}
\eeq
Therefore, if one knows exactly the one-point resolvent for arbitrary
potential, all multi-point resolvents can be calculated by induction.
In the above normalization, the genus expansion has the form
\beq
W(\lambda_1,\dots,\lambda_s)=\sum_{h=0}^{\infty}
\hbar^{2h}
W_{h}(\lambda_1,\dots,\lambda_s),\quad s\geq1,
\label{4.7}
\eeq
which is analogous to genus expansion \rf{Xap2.2}.

The first in the chain of the loop equations \cite{48} of the 1MM is~\cite{Virasoro}
\beq
\oint_{{\cal C}_{\cal D}}\frac{d\lambda}{2\pi
i}\frac{V'(\lambda)}{x-\lambda}W(\lambda)\equiv
\widehat{K}W(x)=
W(x)^2+
\hbar^2
W(x,x)
\label{4.8}
\eeq
where the linear integral operator $\widehat{K}$,
\beq
\widehat{K}f(x)\equiv\oint_{{\cal C}_{\cal D}}\frac{d\lambda}{2\pi i}
\frac{V'(\lambda)}{x-\lambda}f(\lambda)=\left[V'(x)f(x)\right]_-
\label{4.12}
\eeq
projects onto the negative powers\footnote{In order to prove it, one
suffices to deform the integration contour to infinity to obtain
$$
\oint_{{\cal C}_{\cal D}}\frac{d\lambda}{2\pi i}
\frac{V'(\lambda)}{x-\lambda}f(\lambda)=V'(x)f(x)-\left[V'(x)f(x)\right]_+=
\left[V'(x)f(x)\right]_-
$$
} of $\lambda$.
Hereafter, ${\cal C}_{\cal D}$~is a contour encircling all singular points
of $W(\lambda)$, but not the point
$\lambda=p$.
Using Eq.~\theTag{4.5}, one can express
the second term in the r.h.s.\ of loop equation~\theTag{4.8} through
$W(p)$, and Eq.~\theTag{4.8} becomes an equation for
the one-point resolvent \rf{4.2}.

Substituting the genus expansion~\theTag{4.7} in Eq.~\theTag{4.8}, one finds
that $W_h(\lambda)$ for $h\geq1$ satisfy the equation
\beq
\left(\widehat{K}-2W_{0}(\lambda)\right)W_h(\lambda)=\sum_{h'=1}^{h-1}
W_{h'}(\lambda)W_{h-h'}(\lambda)+\frac{\partial }{\partial V(\lambda)}W_{h-1}(\lambda),
\label{4.11}
\eeq
In Eq.~\theTag{4.11}, $W_h(\lambda)$ is expressed through only the
$W_{h_i}(\lambda)$ for which $h_i<h$. This fact allows one to
develop the iterative procedure.

To this end, one needs to use the asymptotics condition
(which follows from the definition of the matrix
integral)
\be
W_h(\lambda)|_{\lambda\to\infty} = \frac{t_0}{\lambda}\delta_{h,0}+O({1}/{\lambda^2}),
\label{Winf}
\ee
and manifestly solve (\ref{4.8}) for genus zero. Then, one could
iteratively find $W_h(\lambda)$ and then restore the corresponding contributions
into the free energy by integration, since
\be
\label{total}
W_h(\lambda)=\dV {\cal F}_h,\quad h\ge 1.
\ee

\subsection{Solution in genus zero} \label{basic}

In genus zero, the loop equation (\ref{4.8}) reduces to
\be
\cI \frac{V^{\prime}(\lambda)}{x-\lambda} W_0(\lambda)=
\left[V'(x)f(x)\right]_- = (W_0(x))^2 \label{plan}
\ee

In order to solve this equation for the planar one-point resolvent
$W_0(p)$, one suffices to note that
\be
\left[V'(\lambda)W_0(\lambda)\right]_-=V'(\lambda)W_0(\lambda)-\left[V'(\lambda)W_0(\lambda)\right]_+
\ee
and, due to (\ref{Winf}), the last term in the r.h.s. is a polynomial
of degree $m-1$, $m$ being the degree of $V'(\lambda)$,
\be
\label{polP}
P_{m-1}(\lambda)  = -\left[V'(\lambda)W_0(\lambda)\right]_+=
-\oint_{\cal C_{\infty}}\frac{dx}{2\pi i}
      \frac{\Vp(x)}{\lambda-x}W_0(x)
\ee
It can be also rewritten as action of the linear operator $\hat r_V
(\lambda)$ acting
to the free energy,
\be\label{R}
P_{m-1}(\lambda)=-\hat r_V(\lambda){\cal F}_0,\ \ \ \ \hat
r_V(\lambda)\equiv\sum_{k,l}(k+l+2)t_{k+l+2}\lambda^k{\partial
\over\partial t_l}
\ee

Then, the solution to \rf{plan}
is
\be
W_0(\lambda) = \frac{1}{2}\Vp(\lambda) - \frac{1}{2}\sqrt{\Vp(\lambda)^2+4P_{m-1}(\lambda)},
 \label{*loop2*}
\ee
where the minus sign is chosen in order to fulfill the asymptotics (\ref{Winf}).
For the polynomial potential of power $m+1$, the resolvent $W_0(\lambda)$ is
a function on complex plane with $m$ cuts, or on a hyperelliptic curve
\be
\label{1mamocu}
y^2 = \Vp(\lambda)^2+4P_{m-1}(\lambda)
\ee
of
genus $g=m-1$, conveniently presented introducing new variable $y$ by
\be
W_0(\lambda) = \frac{1}{2}\left( \Vp(\lambda)-y\right), \label{ansatz}
 \label{*loop3*}
\ee
For generic potential $V(\lambda)$ with $m\to\infty$, curve \rf{1mamocu}
may have an infinite
genus, but we can still consider solutions with only finite number $n$ of cuts
and separate the smooth part of curve \rf{1mamocu} introducing
\be
\label{ty}
y\equiv M(\lambda)\ty, \quad \hbox{and ``reduced" Riemann surface} \quad
{\ty}^2\equiv\prod\nolimits_{\alpha=1}^{2n}(\lambda-\mu_\alpha)
\ee
with all $\mu_\alpha$ distinct.
In what follows, we still assume
$M(\lambda)$ to be a polynomial of degree $m-n$, keeping in mind that $n$ is always
finite and fixed, while $m\geq n$ can be chosen arbitrarily large.
By convention, we set $\ty|_{\lambda\to\infty}\sim \lambda^{n}$, and
$M(\lambda)$ is then\footnote{
Since, due to (\ref{1mamocu}), at large $\lambda$
\be\label{t3}
V'(\lambda)= y(\lambda) + {2t_0\over \lambda}+O(1/\lambda^2)
\ee
}
\be
M(\lambda) = \oint_{\cal C_{\infty}} \frac{dx}{2\pi i}
\frac{\Vp(x)}{(x-\lambda)\ty(x)}
\equiv {\cal M}\prod_i^{m-n}\left(\lambda-\lambda_i\right)
\label{M}
\ee
Note that the values of $M(\lambda)$ at branching points, $M_{m-n}(\mu_\alpha)$
coincide
with the {\it first moments\/} $M_{\alpha}^{(1)}$ of the general
matrix model potential (see~\cite{ACKM,Ak96,KMT}), while the higher moments
are just derivatives at these points. Indeed, by their definition, these moments are
\be
\label{M1}
M_{\alpha}^{(p)}=\oint_{{\cal C}_{\cal D}}\frac{dx}{2\pi i}\frac{V'(x)}
{(x-\mu_\alpha)^p\tilde y(x)},\quad p\ge1
\ee
Then, one deforms the integration contour to
infinity assuming $V'(x)$ is an entire function and, as in (\ref{M}),
make use of the asymptotic conditions
(\ref{t3}), to replace $V'(x)$ in the numerator of (\ref{M1})
by $y(x)=M_{m-n}(x)\tilde y(x)$ subsequently evaluating the integral
by taking the residue at the point $x=\mu_\alpha$ and obtaining
\be
\label{M2}
M_{\alpha}^{(p)}=
\frac{1}{(p-1)!}\,\left.\left(\frac{\d^{p-1}}{\d\lambda^{p-1}}
M(\lambda)\right)\right|_{\lambda=\mu_\alpha}, \ \hbox{e.g.,}\
M_{\alpha}^{(1)}=M(\mu_\alpha).
\ee

Inserting the solution (\ref{ty}), (\ref{M}) into (\ref{ansatz}) and deforming
the contour, one obtains the planar one-point resolvent
with an $n$-cut structure,
\be
W_0(\lambda) = \frac{1}{2}\oint_{{\cal C}_{\cal D}}\frac{dx}{2\pi i} \frac{\Vp(x)}
{\lambda-x}\frac{\ty(\lambda)}{\ty(x)},
\quad \lambda\not\in{\cal D}.
\label{W0}
\ee
The contour ${\cal C_{\cal D}}$ of integration here
encircles the finite number~$n$ of disjoint intervals
\be
{\cal D} \equiv \bigcup_{i=1}^n [\mu_{2i-1},\mu_{2i}],
\quad \mu_1< \mu_2< \ldots < \mu_{2n}.
\ee

Let us now discuss how many free parameters we have in our solution. If one
does not keep genus of the curve fixed, it is given for a generic potential
$V(\lambda)$ by $m+1$ times $t_k$, coefficients of $V'(\lambda)$, and $m-1$ coefficients
$p_k$ of the
polynomial $P_{m-1}(\lambda)$, its leading coefficient being related to $t_0$.
Indeed, using (\ref{Winf}) and (\ref{*loop2*}), one obtains
\be
\left.W_0(\lambda)\right|_{\lambda\to\infty}=
-{P_{m-1}(\lambda)\over V'(\lambda)}+\ldots=-{p_{m-1}\over
(m+1)t_{m+1}}{1\over \lambda}+O({1}/{\lambda^2})=\frac{t_0}{\lambda}+O({1}/{\lambda^2})
\ee
Therefore, totally one has $2m+1$ parameters (including $t_0$). In fact, we usually
fix the leading coefficient of the potential $V(\lambda)$ to be $1/m$ (see s.3),
i.e. ${\cal M}=1$ in (\ref{M}), which leaves us with $2m$ parameters.

If, however, now one fixes the curve of genus $g=n-1$, (\ref{ty}), this
imposes $m-n$ conditions of double points (coinciding branching
cuts)\footnote{One can also give these conditions via different integral
conditions. Say, one can get from (\ref{Winf}) and (\ref{W0}),
the asymptotic conditions
\be
t_0\delta_{k,n}  =  \frac{1}{2}
 \cI \frac{\lambda^k V^{\prime}(\lambda)}{\ty}, \quad k=0,\ldots,n.
 \label{Rand1}
\ee
These conditions are identically satisfied for $m=n$, since $V'(\lambda)\sim
y(\lambda)+O(1/\lambda)=\tilde y(\lambda)+O(1/\lambda)$, while for $m>n$ they impose constraints.
}.
Therefore, one then has $2m-(m-n)=m+n$ parameters. Of these, still $m+1$
parameters are $t_k$, which are variables in the loop equations, while $n-1$
parameters give the arbitrariness of solutions to these loop equations
(possible different choices of the polynomial $P_{m-1}$).

One can arbitrary choose coordinates on this $n-1$-dimensional space of parameters.
It turns out, however, that there is a distinguished set of $n$ independent variables that
parameterize solutions to the loop equations \cite{David,DV},
\be
\label{Sfr}
S_i =
\oint_{A_i}\frac{d\lambda}{4\pi i}\,y=
\oint_{A_i}\frac{d\lambda}{4\pi i}M(\lambda)\ty,
\ee
where $A_i$, $i=1,\dots,n-1$ is the basis of $A$-cycles on the reduced hyperelliptic
Riemann surface \rf{ty} (we may conveniently choose them to be the first $n-1$
cuts) see Fig.1.

Besides canonically conjugated $A$- and $B$-cycles, we also use the linear
combination of $B$-cycles: $\bar B_i\equiv B_i-B_{i+1}$, $\bar B_{n-1}\equiv
B_{n-1}$. Therefore, $\bar B$-cycles encircle the nearest ends of two
neighbor cuts, while all $B$-cycles goes from a given right end of the cut
to the last, $n$th cut. For the sake of definiteness, we order all points
$\mu_\alpha$ in accordance with their index so that $\mu_\alpha$ is to the right of
$\mu_\beta$ if $\alpha>\beta$.

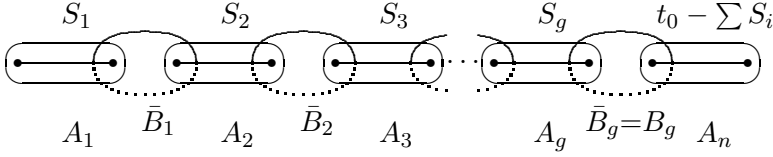
\begin{figure}[h]\label{fig9.1}
\begin{picture}(190,55)(10,10)
\multiput(40,40)(40,0){5}{\oval(30,10)}
\multiput(28,40)(40,0){5}{\line(1,0){24}}
\multiput(28,40)(40,0){5}{\circle*{2}}
\multiput(52,40)(40,0){5}{\circle*{2}}
\thinlines
\qbezier(47,40)(47,48)(60,48)
\qbezier(60,48)(73,48)(73,40)
\thicklines
\qbezier[10](47,40)(47,32)(60,32)
\qbezier[10](60,32)(73,32)(73,40)
\thinlines
\qbezier(87,40)(87,48)(100,48)
\qbezier(100,48)(113,48)(113,40)
\thicklines
\qbezier[10](87,40)(87,32)(100,32)
\qbezier[10](100,32)(113,32)(113,40)
\thinlines
\qbezier(127,40)(127,46)(136,47)
\qbezier(144,47)(153,46)(153,40)
\thicklines
\qbezier[7](127,40)(127,34)(136,33)
\qbezier[7](144,33)(153,34)(153,40)
\thinlines
\qbezier(167,40)(167,48)(180,48)
\qbezier(180,48)(193,48)(193,40)
\thicklines
\qbezier[10](167,40)(167,32)(180,32)
\qbezier[10](180,32)(193,32)(193,40)
\put(39,20){$A_1$}
\put(59,23){$\bar B_1$}
\put(39,50){$S_1$}
\put(79,20){$A_2$}
\put(99,23){$\bar B_2$}
\put(79,50){$S_2$}
\put(119,20){$A_3$}
\put(119,50){$S_3$}
\put(136,40){$\dots$}
\put(158,20){$A_g$}
\put(171,23){$\bar B_g{=}B_g$}
\put(159,50){$S_g$}
\put(199,20){$A_n$}
\put(189,50){$t_0-\sum S_i$}
%
\end{picture}
\caption{Structure of cuts and contours for the reduced Riemann surface.}
\end{figure}

\subsection{Matrix eigenvalue picture: a detour}

The variables $S_i$ can be formally determined as eigenvalues of a differential
operator in $t_k$'s \cite{AMM3}. However, they can be also more
``physically" interpreted in the quasi-classical picture of matrix eigenvalues
(Coulomb gas) when their number (and, therefore, size of the matrix) goes to
infinity. We come now to this interpretation.

To this end, let us first introduce the averaged eigenvalue distribution
\be\label{rho}
\rho(\lambda)\equiv {t_0\over N}\sum_i^N \left< \delta (\lambda-x_i)\right>=
\frac{1}{2\pi i}\lim_{\epsilon\to 0}
         \Big( W(\lambda-i\epsilon)-W(\lambda+i\epsilon) \Big)
\ee
where $x_i$'s are eigenvalues of the matrix $X$. In the planar limit,
this quantity becomes
\be
\rho_0(\lambda)=\frac{1}{2\pi i}\lim_{\epsilon\to 0}
         \Big( W_0(\lambda-i\epsilon)-W_0(\lambda+i\epsilon) \Big)
= \frac{1}{2 \pi}\,\hbox{Im\,}y(\lambda)
\ee
and satisfies the
equation\footnote{Indeed, by definition
$$
\oint_{\infty}{W_0(\lambda)\over x-\lambda}d\lambda=0
$$
Now, using (\ref{*loop3*}) and the definition (\ref{rho}) and pulling out
the contour from infinity, one easily comes to this equation.
}
\be\label{rho-eq}
\not\!\!\int_{\cal D}
 {\rho_0(\lambda)\over x-\lambda}d\lambda={1\over 2}V'(x),\ \ \ \ \forall
p\in {\cal D}
\ee
This averaged eigenvalue distribution becomes the distribution of
eigenvalues in the limit when their number goes to infinity. For the
illustrative purposes, let us do the matrix integral, (\ref{Xap2.1})
performing the matrix $X$ in the form of $U\cdot\diag (x_i)\cdot U^{-1}$ with a unitary
matrix $U$ and then first
making the integration over the ``angular" variables $U$. Then, one comes to
\cite{Mehta}
\be\label{varrho}
e^{{\cal F}}\sim \int\prod_i dx_i\prod_{i>j}(x_i-x_j)^2
e^{-{1\over\hbar}\sum_i V(x_i)}=
$$
$$
=\int\prod_i dx_i e^{-{1\over \hbar^2}
\left(\int V(\lambda)\varrho(\lambda)
- \int \varrho (\lambda)\varrho(\lambda')\log |\lambda-\lambda'|
d\lambda d\lambda '\right)}\equiv \int\prod_i dx_ie^{{1\over
\hbar^2}S_{eff}}
\ee
where we introduced the eigenvalue distribution
\be
\varrho (\lambda)\equiv {t_0\over N}\sum_i\delta(\lambda-\lambda_i)
\ee
Now, in the limit of large $N$, one can use the saddle point approximation
to obtain the equation for $\varrho (\lambda)$. However, one also need to take
into account the constraint
\be
\int\varrho(\lambda)d\lambda=t_0
\ee
by adding to $S_{eff}$, (\ref{varrho}) the Lagrange multiplier term $\Pi_0 (\int\varrho
-t_0)$. Note also that $\varrho(\lambda)$ is a non-negative density. This finally
leads to the saddle point equation
\be\label{varrho-eq}
2\int \varrho(\lambda) \log|x-\lambda|d\lambda=V(x) +\Pi_0,\ \ \ \ \forall
x\in \hbox{support of }\varrho
\ee
The derivative of this equation w.r.t. $p$ coincides with (\ref{rho-eq}).

The variable $t_0$ plays role of the (normalized) total number of eigenvalues,
\be
\label{t0}
t_0 = \frac1{4\pi i} \oint_{\cal C_{\cal D}}yd\lambda =
-\frac{1}2\res_\infty(yd\lambda)
\ee
and the support of $\varrho$ is ${\cal D}$ that consists of $N$ segments ${\cal D}_i$.
Now, following \cite{David,DV}, one may fix the (occupation) numbers of eigenvalues in
each of the segments, $S_i$ (\ref{Sfr}), $i=1,\dots,n-1$. We assume the
occupation number for the last, $n$th cut to be $t_0-\sum_{i=1}^{n-1}S_i\equiv S_n$.
\footnote{It is sometimes convenient to consider $S_n$ instead of $t_0$ as a canonical variable.
However, in all instants we use $S_n$, we specially indicate it for not confusing $S_n$ with the ``genuine''
filling fraction variables $S_i$, $i=1,\dots,n-1$.}
(Obviously, no new parameters $S_i$ arise in the one-cut case.)
We formally attain this by introducing the corresponding chemical potentials
(Lagrange multipliers) $\Pi_i$, \
$i=1,\dots, n-1$, in the variational problem for the free energy,
which therefore becomes in the planar limit
\bea
{S}_{eff} \left[\varrho;S_i,t_0,t_k\right] &=&
-\int_{\cal D} V(\lambda)\varrho(\lambda)d\lambda +
\int\!\!\int_{\cal D} \varrho(\lambda)
\log\left|\lambda-\lambda '\right|
\varrho (\lambda ')d\lambda d\lambda ' +
\nonumber
\\
&&-\Pi_0 \left(\int_{\cal D}\varrho(\lambda)d\lambda - t_0\right)
-\sum_{i=1}^{n-1} \Pi_i \left(\int_{D_i}\varrho(\lambda)d\lambda - S_i\right).
\label{variF}
\eea
while the saddle point equation becomes
\be\label{varrho-eq'}
2\int \varrho(\lambda) \log|x-\lambda|d\lambda=V(x)+\Pi_i +\Pi_0,\ \ \ \ \forall
x\in {\cal D}_i
\ee
and its derivative still coincides with (\ref{rho-eq}).

Therefore, with generic values of the constants $\Pi_i$,
$\varrho_c(\lambda)$ gives the general solution to (\ref{rho-eq}) (or the
planar limit of the loop equation): these constants describe
the freedom one has when solving the loop equation\footnote{Note that, instead
of fixing the occupation numbers, one could use other ways to fix a
solution to the loop equations, see \cite{AMM,AMM2,AMM3}.
}. However, in the matrix model
integral (where there are no any chemical potentials)
one would further vary ${\cal F}_0$ w.r.t. $\Pi_i$ to find the
``true" minimum of the eigenvalue configuration,
\be\label{oc}
{\d {\cal F}_0\over \d S_i}=0,\ \ \ \ \forall i
\ee
This is a set of equation that fixes concrete values of $S_i$ and $\Pi_i$ in
the matrix integral.

Let us now calculate the derivative of ${\cal F}_0$ (\ref{variF}) w.r.t. $S_i$. From
(\ref{variF}), one has
\be
\left.\frac{\partial S_{eff}}{\partial S_i}\right|_{\varrho=\rho_0}=
-\int_{\cal D}d\lambda \frac{\partial \rho_0(\lambda)}{\partial S_i}
\left(V(\lambda)-2\int_{\cal D}d\lambda'\log(\lambda-\lambda')\rho_0(\lambda')\right)
\label{*b2*}
\ee
The expression in the brackets on the r.h.s. of~(\ref{*b2*}) is almost
a variation of (\ref{variF}) w.r.t. the eigenvalue density, which is
\bea
0=\left.\frac{\delta S_{eff}}{\delta \rho(\lambda)}\right|_{\varrho=\rho_0}&=& V(\lambda)-2\int_{\cal
D}d\lambda'\log(\lambda-\lambda')\rho_0(\lambda')+\Pi_i+\Pi_0 \nonumber
\\
&&\quad\hbox{for}\ \lambda\in D_i\subset{\cal D}.
\eea
It is therefore a step
function, $h(\lambda)$ which is constant equal to $\zeta_i\equiv-\Pi_0-\Pi_i$
on each cut $A_i$. One then has
\bea
\frac{\partial {\cal F}_0}{\partial S_i}=
\left.\frac{\partial S_{eff}[\varrho]}{\partial S_i}\right|_{\varrho=\rho_0}=-\int_{\cal
D}d\lambda
\frac{\partial \rho_0(\lambda)}{\partial S_i}h(\lambda)=
-\frac{1}{4\pi i}\sum_{j=1}^n\zeta_j{\partial\over\partial S_i}\oint_{A_j}y(\lambda)d\lambda=\nonumber\\
=-\sum_{j=1}^n\zeta_j{\partial S_j\over\partial S_i}=-\zeta_i+\zeta_n=\Pi_i.
\label{*b4*}\label{Dvdual}
\eea
In particular,
\be\label{Pi0}
{\d{\cal F}_0\over\d t_0}=\Pi_0
\ee

In \cite{JU90} it was proved that
the difference of values of $\Pi_i$
on two neighbour cuts is equal to
\footnote{The simplest way to prove it is to define function $h(\lambda)$ outside the cuts: $h(\lambda)=V(\lambda)-2\int_{\cal
D}d\lambda'\log({\lambda-\lambda'})\rho_0(\lambda')$ and note that $\left.h'(\lambda)\right|_{\lambda\notin {\cal D}}=2W_0(\lambda)$.}
\be
\label{xi}
\zeta_{i+1}-\zeta_i=2\int_{\mu_{2i}}^{\mu_{2i+1}}W_0(\lambda)d\lambda
\ee
i.e.
\be\label{Dvdual1}
\Pi_i=(\zeta_{i+1}-\zeta_i)+(\zeta_{i+2}-\zeta_{i+1})+\ldots+(\zeta_{n-1}-\zeta_{n-2})+(\zeta_n-\zeta_{n-1})
=\oint_{\bar B_i\cup \bar B_{i+1}\cup\dots\cup \bar B_g}yd\lambda
=\oint_{B_i}yd\lambda
\ee

Note that one can calculate
the planar limit free energy that can be obtained
via substituting the saddle point solution $\varrho$ into
(\ref{variF}) is
\be\label{FS}
{\cal F}_0=S_{eff}\left[\varrho_c\right]=
-{1\over 2}\int_{\cal D} V(\lambda)\varrho_c(\lambda)d\lambda
+{1\over 2}\Pi_0 t_0 +{1\over 2}\sum_{i=1}^{n-1} \Pi_i S_i
\ee

In the paper, we choose the solution to the loop equation with fixed
occupation numbers, (\ref{Sfr}). Note that fixing the chemical potentials
(\ref{Dvdual})-(\ref{Dvdual1}) instead, (\ref{oc}),
corresponds just to interchanging $A$- and $B$-cycles
on the Riemann surface (\ref{1mamocu}). However, ${\cal F}_0$ is {\em not} modular
invariant.
Under the change of homology basis, ${\cal F}_0$ transforms in accordance with the
duality transformations \cite{dWM} (which is a particular case of behaviour of
${\cal F}_0$ under the general transformations, \cite{MiMo}).
The higher-genus corrections become also basis-dependent: choosing
$S_i$ or $\Pi_i$ as independent variables, one obtains
different expressions, say, for the genus-one free energy, see s.4.3.

In the next two subsections we are going to demonstrate that the planar loop equation
solution with fixed occupation numbers corresponds to a
Seiberg-Witten-Whitham system.

\subsection{Seiberg--Witten-Whitham theory \label{ss:SW}}

\paragraph{Seiberg-Witten system.}

We call the SW system \cite{WS1}\footnote{Various properties of such systems can
be found in \cite{Mbook,SWbook,GM}.} the following set of data:

\begin{itemize}

\item a family ${\cal M}$ of Riemann surfaces (complex curves) ${\cal C}$
so that the dimension of moduli space of this family coincides with the
genus\footnote{More generally, this can be extended to the curves with auxiliary
involution, or certain directions in moduli space can be "frozen" in a different
way, certain examples can be found in \cite{wdvvlong}.};

\item a meromorphic differential $dS$ whose variations w.r.t. moduli of
curves are holomorphic (this implies existence of a connection on moduli space,
so that this statement has a strict sense, see e.g. \cite{Mbook,BraMa}).
\end{itemize}

These data allow to define a SW {\it prepotential}~\cite{WS1}
related to an integrable system \cite{Kri1,WSI10,WSI1,Mbook,GM}.
First, one introduces the variables (whose number coincides with the genus of ${\cal C}$)
\be
\label{SWa}
S_i\equiv\frac1{4\pi i}\oint_{A_i}dS
\ee
where $A_i$ are $A$-cycles on ${\cal C}$.
As soon as $\frac{\partial dS}{\partial S_i}$ is holomorphic,
from the definition (\ref{SWa}) of $S_i$ and the obvious relation
$\partial S_j/\partial S_i=\delta_{ji}$, one finds that
\be
\frac1{4\pi i}\oint_{A_j}\frac{\partial dS}{\partial S_i}=\delta_{ji},
\ee
i.e.,
\be
\label{canoDV}
{\d dS\over\d S_i} = d\omega_i
\ee
where $d\omega_i$ are the canonically normalized holomorphic 1-differentials,
\be\label{canoDV2}
\frac1{4\pi i}\oint_{A_i}d\omega_j = \delta_{ij}
\ee
Introducing $B$-cycles conjugated
to $A$-cycles: $A_i\circ B_j=\delta_{ij}$, where $\circ$ means intersection form,
one obtains that
\be
{\partial\over\partial S_i}\oint_{B_j}dS=\oint_{B_i}d\omega_j=T_{ij}
\ee
is the period matrix of ${\cal C}$ and is therefore symmetric\footnote{This
follows from the Riemann bilinear relations for
canonical holomorphic differentials (\ref{canoDV})
\bea
0=\int_{\Sigma_g} d\omega_i\wedge d\omega_j&=&
\sum_k\left( \oint_{A_k}d\omega_i\oint_{B_k}
d\omega_j-\oint_{A_k}
d\omega_j\oint_{B_k}d\omega_i \right)=
\nonumber
\\
&=& 4\pi i(T_{ij} - T_{ji})
\label{sypema}
\eea
}. Hence,
there exists a locally defined function $F$ such that
\be
\label{SWF}
{\partial F\over\partial S_i}=\oint_{B_i}dS.
\ee
and this function is called a prepotential.

\paragraph{Generalized Whitham system.}

We now extend the Seiberg-Witten system to the quasiclassical or generalized Whitham
system \cite{Kri1} by introducing extra parameters or times $t_k$ into the game,
we do it mostly following
\cite{RG,MarMir98}\footnote{One can introduce these extra parameters in
the context of supersymmetric SW gauge theories, even without reference to an
integrable system {\em a priori}, see \cite{LNS} and references therein.}.
In order to construct a Whitham
system, one needs to add to the SW data a set of jets of local
coordinates in the vicinity of punctures on ${\cal C}$. In particular case
of \rf{1mamocu}, \rf{ty} and these points (the singularities of $yd\lambda$ coincide
with two $\lambda$-infinities, where we choose the local parameter
$\xi={1\over \lambda}$. We then
introduce a set of {\it meromorphic\/} differentials $d\Omega_k$ with the poles
only at these punctures (as the hyperelliptic curve \rf{1mamocu}, (\ref{ty}) is invariant
w.r.t.  the involution $y\to -y$, from now on we just work with either of the
two infinities, see \cite{RG,MarMir98}) with the behavior
\be\label{*Omega*}
d\Omega_k=\pm {k
\over 2}\left(\xi^{-k-1}+O(1)\right)d\xi,\ \hbox{for} \ \xi\to 0,\ k>0
\ee
where signs are different for the different infinities.
We also introduce the bipole or 3rd-kind Abelian
differential, with two simple poles, located at two infinities:
\be\label{bip}
\res_{\infty}d\Omega_0=-\res_{\infty_-}d\Omega_0=-1
\ee
Then, the generalized Whitham system is generated by a set of equations on these
differentials and on the holomorphic differentials $d\omega_i$:
\be\label{*We*}
{\partial d\Omega_p\over\partial t_k}={\partial d\Omega_k\over\partial t_p},
\qquad 2{\partial d\Omega_k\over\partial S_i}=
{\partial d\omega_i\over\partial t_k},\qquad
{\partial d\omega_i\over\partial S_j}={\partial d\omega_j\over\partial S_i}
\ee
where the partial derivatives are supposed to be taken at constant hyperelliptic
co-ordinate $\lambda$.

These equations imply that there exists a differential $dS$
such that (see also (\ref{canoDV})
\be\label{ds}
{\partial dS\over\partial S_i}=d\omega_i,\qquad
{\partial dS\over\partial t_k}=2d\Omega_k
\ee
Note, however, that the meromorphic differentials (\ref{*Omega*}),
(\ref{bip}) are defined up to linear combinations of the holomorphic
differentials. Since we consider $S_i$ and $t_k$ to be independent
variables,
this ambiguity is removed merely by imposing the condition \cite{RG,MarMir98}
\be
\label{I}
\frac{\partial S_i}{\partial t_k}=\frac1{2\pi i}\oint_{A_i}d\Omega_k=0 ~\forall \ i,k
\ee
Now we can invariantly introduce variables $t_k$ via the
relations\footnote{In what follows,
we call the $\infty$-point the point
$\infty_+$, or $\lambda=\infty$ on the ``upper" (physical) sheet of hyperelliptic Riemann
surface (\ref{ty}) corresponding to the positive sign of the square root.}
\be
\label{LGtimes}
t_k = -{1\over k}\res_\infty\left(\lambda^{-k}dS\right),
\quad
k=1,\dots,m,\ \ \ t_0=\res_{\infty}dS\ \ \hbox{(cf. (\ref{t0}))}
\ee
Then, one defines the prepotential that depends on both $S_i$ and $t_k$ via the old
relation (\ref{SWF}) and the similar relations
\be
\label{v41}
{\partial F\over \partial t_k}=\frac12\res_{\lambda=\infty}(\lambda^{k}dS)\equiv \frac12v_k,\quad k=1,\dots,m.
\ee
\be\label{dut0}
{\d F\over\d t_0}=\int^{\infty_+}_{\infty_-}dS
\ee
The latter integral, which is naively divergent, is
still to be supplemented with some proper regularization, we discuss this in
detail in the next subsection.

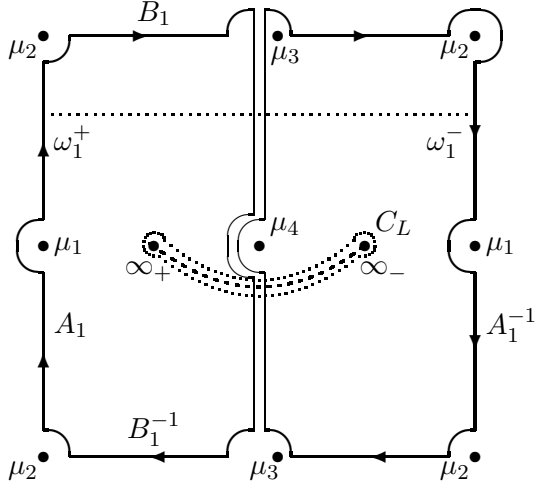
\begin{figure}[tp]
\setlength{\unitlength}{0.7mm}%
\begin{picture}(0,100)(-85,-50)
\thicklines
\put(-41,5){\line(0, 1){30}}
\put(-41,-5){\line(0, -1){30}}
\put(41,5){\line(0, 1){30}}
\put(41,-5){\line(0, -1){30}}
\put(-36,40){\line(1, 0){30}}
\put(6,40){\line(1, 0){30}}
\put(-36,-40){\line(1, 0){30}}
\put(6,-40){\line(1, 0){30}}
\put(-41,0){\oval(10,10)[l]}
\put(41,0){\oval(10,10)[l]}
\put(-41,40){\oval(10,10)[br]}
\put(-41,-40){\oval(10,10)[tr]}
\put(41,-40){\oval(10,10)[tl]}
\put(41,40){\oval(10,10)[br]}
\put(41,40){\oval(10,10)[t]}
\put(-1,40){\oval(10,10)[tl]}
\put(1,40){\oval(10,10)[tr]}
\put(-1,-40){\oval(10,10)[tl]}
\put(1,-40){\oval(10,10)[tr]}
\put(-41,20){\vector(0,1){0}}
\put(-41,-20){\vector(0,1){0}}
\put(41,20){\vector(0,-1){0}}
\put(41,-20){\vector(0,-1){0}}
\put(-21,40){\vector(1,0){0}}
\put(21,40){\vector(1,0){0}}
\put(-21,-40){\vector(-1,0){0}}
\put(21,-40){\vector(-1,0){0}}
\put(-41,40){\circle*{2}}
\put(-41,-40){\circle*{2}}
\put(41,-40){\circle*{2}}
\put(41,40){\circle*{2}}
\put(-41,0){\circle*{2}}
\put(41,0){\circle*{2}}
\put(0,0){\circle*{2}}
\put(3.5,40){\circle*{2}}
\put(3.5,-40){\circle*{2}}
\put(-20,0){\circle*{2}}
\put(20,0){\circle*{2}}
\qbezier[30](-18,1)(0,-14)(18,1)
\qbezier[30](-19,-2)(0,-17)(19,-2)
\qbezier[20](-20,0)(0,-15.5)(20,0)
\qbezier[20](-20.15,0.1)(0,-15.5)(19.85,-0.1)
\qbezier[20](-20.3,0.2)(0,-15.5)(19.7,-0.2)
\qbezier[20](-20.45,0.3)(0,-15.5)(19.55,-0.3)
\qbezier[20](-20.6,0.4)(0,-15.5)(19.4,-0.4)
\qbezier[20](-20.75,0.5)(0,-15.5)(19.25,-0.5)
\qbezier[50](-41,25)(0,25)(41,25)
\qbezier[4](-18,1)(-18,2.5)(-20,2.5)
\qbezier[5](-19,-2)(-22,-2)(-22,0)
\qbezier[5](-20,2.5)(-22,2.5)(-22,0)
\qbezier[4](18,1)(18,2.5)(20,2.5)
\qbezier[5](19,-2)(22,-2)(22,0)
\qbezier[5](20,2.5)(22,2.5)(22,0)
\put(43,0){\makebox(0,0)[lc]{$\mu_1$}}
\put(-39,0){\makebox(0,0)[lc]{$\mu_1$}}
\put(-42,-41){\makebox(0,0)[rt]{$\mu_2$}}
\put(-42,39){\makebox(0,0)[rt]{$\mu_2$}}
\put(40,39){\makebox(0,0)[rt]{$\mu_2$}}
\put(40,-41){\makebox(0,0)[rt]{$\mu_2$}}
\put(5,38){\makebox(0,0)[ct]{$\mu_3$}}
\put(1,-41){\makebox(0,0)[ct]{$\mu_3$}}
\put(2,2){\makebox(0,0)[lb]{$\mu_4$}}
\put(-21,-3){\makebox(0,0)[ct]{$\infty_+$}}
\put(19,-3){\makebox(0,0)[lt]{$\infty_-$}}
\put(22,2){\makebox(0,0)[lb]{$C_L$}}
\put(-39,23){\makebox(0,0)[lt]{$\omega_1^+$}}
\put(39,23){\makebox(0,0)[rt]{$\omega_1^-$}}
\put(-20,42){\makebox(0,0)[cb]{$B_1$}}
\put(-20,-38){\makebox(0,0)[cb]{$B_1^{-1}$}}
\put(-39,-15){\makebox(0,0)[lc]{$A_1$}}
\put(43,-15){\makebox(0,0)[lc]{$A_1^{-1}$}}
\thinlines
\put(-1,45){\line(0, -1){39}}
\put(1,45){\line(0, -1){40}}
\put(-1,-35){\line(0, 1){29}}
\put(1,-35){\line(0, 1){30}}
\put(-1,0){\oval(10,12)[l]}
\put(1,0){\oval(10,10)[l]}
\end{picture}
\caption{Cut Riemann surface from Fig.~\ref{fi:cuts}. The integral over the
boundary can be divided into several pieces (see formula (\ref{dpm})). In
the process of computation we use the fact that the boundary values of
Abelian integrals
$\omega_j^\pm$ on two copies of the cut differ by period integral of
the corresponding differential
$d\omega_j$ over the dual cycle. We add the two infinity points and
the additional, $n$th, cut dividing the surface in two sheets. We present
the logarithmic cut between these two points (about which we draw the
standard integration contour $C_L$). Integrals over small circles around
the points $\mu_\alpha$ are relevant only when calculating the
third-order derivatives w.r.t. the canonical variables $t_I$.}
\label{fi:cut}
\end{figure}

In fact, one still needs to prove such a prepotential exists \cite{RG,MarMir98},
by checking the second derivatives are symmetric. This can be verified using
the Riemann bilinear relations,
\bea
{\d v_k\over \d S_i}&=&
2\res_\infty\left(\Omega_kd\omega_i\right) =
\oint_{\d\Sigma_g}\Omega_kd\omega_i =
\nonumber
\\
&=&2\sum_{l=1}^g\left(\int_{B_l}\Omega_k^+d\omega_i - \int_{B_l}\Omega_k^-
d\omega_i\right) -
2\sum_{l=1}^g\left(\int_{A_l}\Omega_k^+d\omega_i - \int_{A_l}\Omega_k^-
d\omega_i\right) =
\nonumber
\\
&=& 2\sum_{l=1}^g \left(\oint_{A_l}d\Omega_k\oint_{B_l}d\omega_i -
\oint_{B_l}d\Omega_k\oint_{A_l}d\omega_i\right) = 2\oint_{B_i}d\Omega_k
={\d \Pi_i\over \d t_k}.
\label{vvS}
\eea
Here $\d\Sigma_g$ is the cut reduced Riemann surface (\ref{ty}) (see
fig.~\ref{fi:cut}), \ $\Omega_k^{\pm}$
are values of $\Omega_k\equiv \int d\Omega_k$ on two sides of the
corresponding cycle (they are related by the integral over the dual
cycle), and we have used (\ref{v31}).

Similarly, for the derivatives w.r.t. the times $t_k$, one has
analogously to (\ref{vvS})
\bea
\frac12\left({\d v_k\over \d t_p}-{\d v_p\over \d t_k}\right)&=&
\res_\infty\left((\Omega_k)_+d\Omega_p-d\Omega_k(\Omega_p)_+\right) =
\nonumber
\\
&=&\res_\infty\left(\Omega_kd\Omega_p\right) =
\oint_{\d\Sigma_g}\Omega_kd\Omega_p =\nonumber
\\
&=& \sum_{l=1}^g \left(\oint_{A_l}d\Omega_k\oint_{B_l}d\Omega_p -
\oint_{B_l}d\Omega_k\oint_{A_l}d\Omega_p\right)\,{=}\,0.
\label{vvv}
\eea
and
\bea
0&=&\int_{\Sigma_g} d\omega_i\wedge d\Omega_0=
\sum_{l=1}^g\left( \oint_{A_l}d\omega_i\oint_{B_l}
d\Omega_0-\oint_{A_l}
d\Omega_0\oint_{B_l}d\omega_i \right) +
\nonumber
\\
&&\quad +\res_\infty(d\omega_i)\int_{\infty_-}^{\infty_+}d\Omega_0
- \res_\infty(d\Omega_0)\int_{\infty_-}^{\infty_+}d\omega_i=
\nonumber
\\
&=& \oint_{B_i}d\Omega_0 - \int_{\infty_-}^{\infty_+}d\omega_i.
\label{syToda}
\eea

\subsection{Free energy as prepotential of SWW system}

We now associate a Seiberg-Witten-Whitham (SWW) system with the planar limit of the matrix-model
free energy.

\paragraph{Matrix integral as a SW system.}

The family ${\cal M}$ in this case is the family of $h=n-1$
reduced Riemann surfaces described by (\ref{1mamocu}) or (\ref{ty}).
In different words, there is no information in ${\cal M}$ about
the additional polynomial $M_{m-n}(\lambda)$, which
is present, however, for \rf{ty} in the differential $dS$.
The role SW differential is played by
\be\label{dS}
dS=yd\lambda
\ee
Consider its variation w.r.t. $S_i$, the variation over moduli ${\cal M}$
does not change the genus of the reduced Riemann surface as
well as the highest degree of the additional polynomial $M_{m-n}(\lambda)$.
Moreover, considering the times of the potential $V'_m(\lambda)$ to be
{\it independent\/} on the parameters $S_i$, we assume $\delta V'/\delta S_i\equiv0$.
Below, by $\delta$ and $\delta_S$ we denote the respective
general variation and variation specifically w.r.t. the moduli parameters $S_i$.

Using (\ref{1mamocu}), (\ref{ty}), one obtains for the
{\it general\/} variation~$\delta dS$:
\footnote{Note that the variation $\delta$ differs nevertheless from loop
insertion (\ref{4.6}) because
the former does not change, by definition, the degree of the
polynomial $M_{m-n}(\lambda)$.}
\bea
\delta dS&=&\delta\left( M_{m-n}(\lambda)
{\tilde y}(\lambda)\right)d\lambda=
\nonumber
\\
&=&{g_{2n}(\lambda)
\delta M_{m-n}(\lambda)+
{1\over 2}M_{m-n}(\lambda)
\delta g_{2n}(\lambda) \over {\tilde y}(\lambda)}d\lambda,
\label{v1}
\eea
where the polynomial expression in the numerator is of maximum
degree $m+n-1$ (since the
highest term of $M_{m-n}$ is fixed).
On the other hand, under~$\delta_S$ which
does not alter the potential, we obtain
from (\ref{1mamocu}), (\ref{ty}) that
\be\label{v2}
\delta_S dS=-{1\over 2}{\delta_S P_{m-1}(\lambda)\over
M_{m-n}(\lambda){\tilde y}(\lambda)}d\lambda.
\ee
Because this variation is just a particular case of (\ref{v1}), we obtain
that zeros of $M_{m-n}(\lambda)$ in the
denominator of (\ref{v2}) must be exactly {\it cancelled} by zeros of the
polynomial $\delta_S P_{m-1}(\lambda)$ in the numerator,
so the maximum degree of the polynomial in
the numerator is $n-2$ (because, again,
the highest-order term of $P_{m-1}(\lambda)$ is fixed by asymptotic
condition (\ref{t3}) and is not altered by variations $\delta_S$).
We then come to the crucial observation that the variation
$\delta_S dS$ is {\it holomorphic} on the curve (\ref{ty}), as it should be
for the SW differential.

The canonical 1-differentials on the reduced
Riemann surface ${\tilde y}(\lambda)$ have the form
\be
{\d dS\over\d S_i}=d\omega_i={ H_i(\lambda)d\lambda\over\tilde y(\lambda)}
\ee
and $ H_{i}(\lambda)$ are
polynomials of degrees at most $n-2$.
The normalization condition (\ref{canoDV2})
unambiguously fixes the form of the polynomials $ H_{i}(\lambda)$.

Comparing now (\ref{Dvdual1}) and (\ref{SWF}), one could indeed identify the planar
limit ${\cal F}_0$ of the 1MM free energy
with an SW prepotential $F$. However, in order to specify this equivalence
further, one needs to work out the $t$-dependence of the free energy, i.e.
consider the generalized Whitham system.

\paragraph{Matrix integral as a Whitham system.}

To this end, let us check that differential (\ref{dS})
on curve (\ref{ty}) with the
relation for moduli (\ref{1mamocu}), (\ref{ty}) does satisfy (\ref{ds}).

Indeed, we have proved the first set of relations (\ref{ds}) in the previous
paragraph. Now let us consider variations of the potential, i.e., variations
w.r.t. Whitham times $t_k$. Then, we obtain instead of (\ref{v2})
\be\label{v3}
\delta dS=-{1\over 2}{\delta \left((V'_m)^2(\lambda)-P_{m-1}(\lambda)\right)
\over M_{m-n}(\lambda){\tilde y}(\lambda)}d\lambda
\ee
while (\ref{v1}) still holds. Repeating the argument of the previous
paragraph, we conclude that the zeroes of $M_{n-k}(\lambda)$ cancel from the
denominator and, therefore, the variation may have pole only at
$\lambda=\infty$, or $\eta=0$, i.e., at the puncture. In order
to find this pole, we
use (\ref{v1}), which implies that
$dS=M_{m-n}(\lambda){\tilde y}(\lambda)d\lambda\to (V'_m(\lambda)
+O({1\over\lambda}))d\lambda$ and, therefore, the variation of $dS$ at large
$\lambda$ is completely determined by the variation of $V'_m(\lambda)$.
Parameterizing $V(\lambda)=\sum^{m+1}_{k=1}t_k{\lambda^{k}}$, we obtain
(\ref{ds}) up to a linear combination of holomorphic differentials. One
may fix the normalization of $d\Omega_k$ that are also defined up to
a linear combination of holomorphic differentials in order to make Eq.~(\ref{ds})
{\it exact}. We already discussed (see \ref{I}) that this normalization and,
therefore, unambiguous way the variables $S_i$ depend on the
coefficients of $P_{m-1}$ is fixed by the condition \cite{RG,MarMir98}
\be
\label{v31}\label{dVSi}
\frac{\partial S_i}{\partial t_k}=\frac1{2\pi i}\oint_{A_i}d\Omega_k=0\ \forall \ i,k,\ \ \
\hbox{or }\ \  {\d S_i\over\d V(\lambda)}=0\ \ i=1,\ldots,n-1
\ee
The derivatives of $dS$ w.r.t. the times are
\bea
2 d\Omega_k \equiv {\d dS \over \d t_k} &=&
{V'(\lambda)k\lambda^{k-1} d\lambda\over y} +
{1\over 2}\sum_{j=0}^{m-2}{\d P_j\over\d t_k}{\lambda^j d\lambda\over y}
\nonumber
\\
&\equiv&{ H_{n+k-1}(\lambda) d\lambda\over \tilde y(\lambda)},
\label{omes}
\eea
and the normalization conditions (\ref{dVSi})
together with the asymptotic expansion
\bea
2 d\Omega_k(\lambda)|_{\lambda\to\infty}&=&{k\lambda^{k-1}d\lambda}
+O(\lambda^{-2})d\lambda=
\nonumber
\\
&=&{k\lambda^{k-1}d\lambda}+\sum_{m=1}^{\infty}
c_{km}\lambda^{-1-m}d\lambda.
\label{ass}
\eea
fixes uniquely the coefficients of the
corresponding polynomials $ H_{n+k-1}$
of degrees $n+k-1$.

Now we again
find that the prepotential $F$ defined in (\ref{v41}) {\it coincides\/}
with ${\cal F}_0$.
To this end, we apply the formula similar to
(\ref{*b2*}) with the only difference that the potential~$V(\lambda)$
itself is changed. We then obtain (see (\ref{*b2*})-(\ref{Dvdual}))
\bea
\frac{\partial {\cal F}_0}{\partial t_k}&=&
-\frac1{4\pi i}\oint_{\cal D}d\lambda
\frac{\partial y(\lambda)}{\partial t_k}\cdot h(\lambda)
-\frac1{4\pi i}\oint_{\cal D}d\lambda y(\lambda) {\lambda^k}=
\nonumber
\\
&=&-\sum_{i=1}^{n-1} \frac{\partial S_i}{\partial t_k}(\zeta_i-\zeta_n)+
\frac12\res_{\lambda=\infty}\lambda^{k}dS,
\label{b22}
\eea
which by virtue of (\ref{v31}) gives (\ref{v41}).

\paragraph{On $t_0$-dependence of prepotential.}

\begin{figure}[tb]
\vskip .2in
\setlength{\unitlength}{0.8mm}%
\begin{picture}(0,40)(15,15)
\thicklines
\curve(60,30, 62,33.5, 65,35, 70,36, 80,36.5, 90,36, 95,35, 98,33.5, 100,30, 98,26.5, 95,25, 90,24, 80,23.5, 70,24, 65,25, 62,26.5, 60,30)
\curve(64,29, 66,31, 68,29, 70,31, 72,29, 74,31, 76,29, 78,31, 80,29, 82,31, 84,29, 86,31, 88,29, 90,31, 92,29, 94,31, 96,29)
\put(80,40){\makebox(0,0)[rb]{$A$}}
\curve(120,30, 122,33.5, 125,35, 130,36, 140,36.5, 150,36, 155,35, 158,33.5, 160,30, 158,26.5, 155,25, 150,24, 140,23.5, 130,24, 125,25, 122,26.5, 120,30)
\curve(124,29, 126,31, 128,29, 130,31, 132,29, 134,31, 136,29, 138,31, 140,29, 142,31, 144,29, 146,31, 148,29, 150,31, 152,29, 154,31, 156,29)
\put(110,40){\makebox(0,0)[rb]{$B$}}
\curve(92,26.5, 90,30, 92,33.5, 95,35, 100,36, 110,36.5, 120,36, 125,35, 128,33.5, 130,30, 128,26.5)
{\curvedashes[1mm]{0,1,2}
\curve(92,33.5, 90,30, 92,26.5, 95,25, 100,24, 110,23.5, 120,24, 125,25, 128,26.5, 130,30, 128,33.5)}
\put(204,35){\circle*{1}}
\put(204,25){\circle*{1}}
{\thinlines
\curvedashes[1mm]{1,3,2}
\curve(204,35, 157,35, 154,34, 153,33, 152,30, 153,27, 154,26, 157,25, 204,25)
\curvedashes[1mm]{5,1}
\curve(204,35, 157,35, 154,34, 153,33, 152,30, 153,27, 154,26, 157,25, 204,25)}
\thinlines
\put(180,37){\makebox(0,0)[cb]{$C_L$}}
\put(207,40){\makebox(0,0)[cb]{$C_\Lambda$}}
\put(208,35){\makebox(0,0)[cc]{${\scriptstyle \infty_+}$}}
\put(208,25){\makebox(0,0)[cc]{${\scriptstyle \infty_-}$}}
\curve(153.5,30, 154,31.8, 154.5,32.8, 155.5,33.5, 157.5,34, 164,34, 199,34, 200,34, 201,33.5, 205,32, 209,32, 210,32.3, 211,32.8, 212,34, 212.3,35)
\curve(150.5,30, 151,32.2, 151.5,33.7, 152.5,35, 154.5,36, 162,36, 199,36, 200,36, 201,36.5, 205,38, 209,38, 210,37.7, 211,37.2, 212,36, 212.3,35)
\curvedashes[0.5mm]{2,1}
\curve(153.5,30, 154,28.2, 154.5,27.2, 155.5,26.5, 157.5,26, 164,26, 199,26, 200,26, 201,26.5, 205,28, 209,28, 210,27.7, 211,27.2, 212,26, 212.3,25)
\curve(150.5,30, 151,27.8, 151.5,26.3, 152.5,25, 154.5,24, 162,24, 199,24, 200,24, 201,23.5, 205,22, 209,22, 210,22.3, 211,22.8, 212,24, 212.3,25)
\end{picture}
\caption{Cuts in the $\lambda$-, or ``eigenvalue," plane for the
planar limit of 1MM. The eigenvalues are supposed to be
located ``on" the cuts. The distribution of eigenvalues is governed by the
period integrals $S_i = \oint_{A_i} \rho(\lambda)d\lambda$ along the
corresponding cycles,
and the dependence of the free energy on ``occupation numbers" $S_i$ is
given by quasiclassical tau-function ${\d{\cal \F_0}\over\d S_i} =
\oint_{B_i} y(\lambda)d\lambda$. We must add the logarithmic
cut between two copies of the infinity on two sheets of the
hyperelliptic Riemann surface in order to calculate the derivative
w.r.t. the variable $t_0$.}
\label{fi:cuts}
\end{figure}
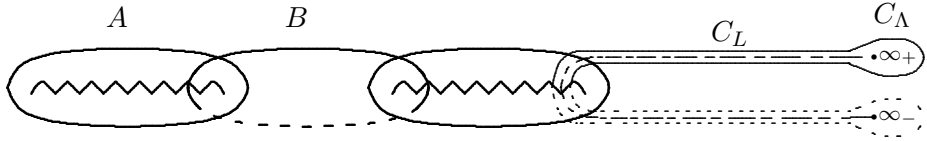

Thus, we have proved the derivatives of the SWW prepotential and of the
matrix model free energy w.r.t. $S_i$'s and $t_k$'s coincide. However, there
is a subtlety of exact definition of integral (\ref{dut0}) for $\d F/\d t_0$.
This quantity is to be compared with the derivative $\d{\cal F}_0/\d t_0$,
(\ref{Pi0}) equal to the integral
$\oint_{\cal D}\log|\lambda-y|dS-V(y)$, where the
reference point $y$ is to be chosen on the last, $n$th, cut, while the
expression itself does not depend on the actual local position of the
reference point. It is convenient to choose it to be
$\mu_{2n}\equiv b_n$---the
rightmost point of the cut. We can then invert the contour
integration over the support $\cal D$ to the integral along the contour
that runs first along the upper side of the logarithmic cut
from $b_n$ to a regularization point $\Lambda$, then
over the circle $C_\Lambda$ of large radius $|\Lambda|$ and then back over
the lower side of the logarithmic cut in the complex plane. In order to
close the contour on the hyperelliptic Riemann surface under consideration,
we must add the integration over the corresponding contour on the {\it second}
sheet of the surface
as shown in Fig.~\ref{fi:cuts}; we
let $C_L$ denote the completed integration contour, and
it is easy to see that such an
additional integration just double the value of the integral.

It is easy to see that all the singularities appearing at the upper
integration limit (i.e., at the point~$\Lambda$) are exactly cancelled
by the contribution coming when
integrating the expression $dS\log(\lambda-b_n)$
along the circle $C_\Lambda$; in fact, the latter can be
easily done, the result is $-2\pi i\bigl(S(\Lambda)-\{S(b_n)\}_+\bigr)$, where the
function $S(\lambda)$ is the (formal)
primitive of $dS$ (which includes the logarithmic term),
and the symbol
$\{\cdot\}_+$ denotes the projection to the strictly polynomial part of the
expression in the brackets. Using the large-$\lambda$ asymptotic expansion
of the differential $dS$,
\be
\label{asympt}
dS(\lambda)|_{\lambda\to\infty}=V'(\lambda)d\lambda
+{t_0\over\lambda}d\lambda+O(\lambda^{-2})d\lambda,
\ee
we obtain that $(S(b_n))_+$ just cancels the term $V(b_n)$,
and we eventually find that
\bea
{\d F\over\d t_0}&=&1/2\left(\oint_{C_L}\log(\lambda-b_n)dS-2V(b_n)\right)=
\nonumber
\\
&=&2\pi i\left(\int_{b_n}^\Lambda dS-S(\Lambda)\right),
\label{dfdt0}
\eea
where $C_L$ is the contour described above (see also Fig.~\ref{fi:cuts}),
which by convention encircles the logarithmic cut between two infinities
on two sheets of the Riemann surface and passes through the last, $n$th, cut.

Thus, we proved that $\d F/\d t_0$ coincides $\d{\cal F}_0/\d t_0$,
(\ref{Pi0}). This completes the proof that the
derivatives of $F$ and ${\cal F}_0$ w.r.t. all $S_i$'s and $t_k$,
$k=0,1,...$ coincide.
Therefore, the planar limit ${\cal F}_0$ of the 1MM free energy
is indeed the SWW prepotential or quasiclassical tau-function. Note that this
identification of the matrix model free energy and the SWW
prepotential is crucially based on formula (\ref{v31}) which fixes
solutions to the loop equations (see ss.1.2-1.3). Moreover, making higher
genera calculations, we shall solve the loop equations with similar
additional constraints that fix the solution, see s.4 and formula
(\ref{A-cycle-F}) below.

We now introduce the (complete) set of canonical variables $\{S_i,\
i=1,\dots, n-1;\ t_0;\ t_k,\ k=1,\dots,m\}$, which we uniformly denote $\{t_I\}$
(in what follows, Latin capitals indicate any quantity from this set).
From (\ref{canoDV}), (\ref{bip}), and (\ref{omes}), we then obtain the
general relation
\be
\label{dOI}
{\d dS\over\d t_I}\equiv d\Omega_I=
{ H_{I}(\lambda)d\lambda\over\tilde y(\lambda)},
\ee
where $ H_{I}(\lambda)$ are polynomials.

Asymptotic formulas (\ref{bip}) and (\ref{ass}) imply
that {\it derivatives\/} of all the quantities $d\Omega_I$
w.r.t. any parameter
are regular at infinity and may have singularities
{\it only\/} at the ramification points $\mu_\alpha$ of reduced
Riemann surface (\ref{ty}).

\section{Second derivatives of free energy \label{s:geometry}}

\subsection{Bergmann bidifferential \label{ss:BPQ}}
In the previous section  we demonstrated how to calculate one-point
resolvent. This required the knowledge of the first derivatives of the
matrix model free energy. These could be expressed through the local
quantities (integrals of differentials) on Riemann surfaces
(and allowed us to associate our construction with the Whitham system).

In this section, we apply this procedure to the two-point
resolvent, which requires the knowledge of the second derivatives.
Here, instead of differentials with some prescribed
properties of holomorphicity, the main object we need is a
{\it bi}differential: the
{\it Bergmann kernel} (canonically normalized bidifferential in Fay's terminology)
which is the bi-differential on a Riemann
surface~$\Sigma_g$ being the double derivative of logarithm of the prime form $E(P,Q)$
such that it is symmetrical in its arguments $P,Q\in \Sigma_g$
and has the only singularity at the coinciding arguments where it has the
behavior (see~\cite{Fay})
\be
B(P,Q)=\left(\frac{1}{(\xi(P)-\xi(Q))^2}+\frac16S_B(P)+o(1)\right)d\xi(P)d\xi(Q),
\label{*Bergmann*}
\ee
in some local coordinate $\xi(P)$ in the vicinity of a point $P\in{\cal C}$;
$S_B(P)$ is the Bergmann projective connection
($S_B(P)$ transforms as a quadratic differential plus the Schwarzian derivative
under an arbitrary variable changing; this
transformation law is the same as for the energy--momentum tensor of the (free) scalar
field, see~\cite{BPZ}). As it stands, we can add to (\ref{*Bergmann*}) any bilinear
combination of Abelian 1-differentials $d\omega_i$; we fix the normalization claiming
vanishing all the integrals over $A$-cycles of $B(P,Q)$:
\be
\oint_{A_i}B(P,Q)=0,\ \hbox{for}\ i=1,\dots,g,
\label{*vanish*}
\ee
and, due to the symmetricity property, the integral may be taken over any of the
variables $P$ or $Q$.

The prime form, a fundamental object on the Riemann surface is
defined as follows.
Consider the Jacobian~$J$, which is a $h$-dimensional torus
defined by the period map of the curve~$\Sigma_g$. Recall that the Abel map
$\Sigma_g\mapsto J:\ P\to \vec x(P)\equiv\left\{\int_{P_0}^Pd\omega_i\right\}$, where $P_0$ is a
reference point, set into the correspondence to each point $P$ of the complex curve
the vector in the Jacobian, and we also introduce the theta function
$\Theta_{[\alpha]}(\vec x)$ of an odd characteristic $[\alpha]$
that becomes zero at $\vec x=0$. Introduce also the "normalizing" holomorphic
half-differentials $h_{\alpha}(P)$ determined for the points of $P\in\Sigma_g$
and characteristics $\alpha$ by
$$
h^2_{\alpha}(P)=\sum_{i=1}^g\frac{\d \Theta_{[\alpha]}(0)}{\d x_i}d\omega_i(P).
$$
The explicit expression for the
prime form $E(P,Q)$ that has a single zero on the Riemann surface~$\Sigma_g$
then reads
\be
E(P,Q)=\frac{\Theta_{[\alpha]}(\vec x(P)-\vec x(Q))}{h_{\alpha}(P)h_{\alpha}(Q)}
\label{*E(P,Q)*}
\ee
while the Bergmann kernel is just
\be\label{BO}
B(P,Q)=d_Pd_Q\log E(P,Q)
\ee
which can be immediately stated by analytical properties and zero
$A$-periods.

Note that, up to a holomorphic part, the Bergmann
bidifferential is nothing but the scalar Green function, see \cite{RG,Knizhnik,LeMo}.
Another useful Green function, that is, the fermionic one, $\Psi_e(P,Q)$
is defined to be a holomorphic
$1/2$-differential in both variables but the point $P=Q$ where it has the first
order pole with unit residue. It also depends on the choice of
theta-characteristics $e$ (boundary conditions for the fermions) and is
manifestly given by
\be\label{Szego}
\Psi_e(P,Q)={\Theta_e (\vec x(P)-\vec x(Q))\over\Theta_e(\vec 0) E(P,Q)}
\ee
The square of the Szeg\"o kernels and the bi-differential $B(P,Q)$
are related by the identity \cite{Fay} (Proposition 2.12; see also
Appendices A,B in \cite{RG,Mbook}):
\be
\left.\Psi_e(P,Q)\Psi_{-e}(P,Q) =  B(P,Q) +
d\omega_i(P)d\omega_j(Q)
\frac{\partial^2}{\partial x_i\partial x_j}
\log \Theta_e(\vec x)\right|_{\vec x=0}
\label{SzegoB}
\ee
This allows one to express $B(P,Q)$ through the
square of the Szeg\"o kernel (note that, for the half-integer characteristics,
$-e$ is equivalent to $e$).

As we shall see in the next subsection, the Bergmann kernel generates the
differentials $d\Omega_k$. Therefore, formula (\ref{BO}) would allow one to
express these latter through the prime form. Similarly, the bipole differential
(\ref{bip}) can be rewritten through the prime form as
\be
\label{bp}
d\Omega_{0} = d\log{E(P,\infty_+)\over E(P,\infty_-)}
\ee
The primitive of differential (\ref{bp}) (which we need in what follows)
then obviously develops the logarithmic cut
between the points of two infinities on the Riemann surface.


\subsection{2-point resolvent
}

Now one can easily express the 2-point resolvent
$W_0(\lambda,\mu)$ in terms of $B(P,Q)$
on hyperelliptic curve (\ref{ty}), where we now
use the hyperelliptic co-ordinate $\lambda=\xi(P)$ and $\mu=\xi(Q)$.
Indeed, let us use (\ref{4.5}), (\ref{v41}) and (\ref{ds}) to
obtain
\bea
W_0(\lambda,\mu)d\mu d\lambda=\sum^{\infty}_{k,l\geq 0}{d\mu d\lambda\over
\mu^{k+1}\lambda^{l+1}} \res_{\infty_+} x^k d\Omega_l(x) =
\sum_{l\geq 0}{d\lambda\over\lambda^{l+1}}d\tilde\Omega_l(\mu)
\label{BPQ}
\eea
where $d\tilde\Omega_k=d\Omega_k-{1\over 2}dx^k$ is the meromorphic second-kind Abelian
differential with the only singularity at $\infty_-$, where it behaves like
\be\label{tildeomega}
d\tilde\Omega_k(\mu)=
-k\left(\mu^{k-1}+O(1)\right)d\mu,\ \hbox{for} \ \mu\to
\infty_-,\ k>0
\ee
and has vanishing $A$-periods $\oint_{A_i}d\tilde\Omega_k=0$.
Therefore, $W_0(\lambda,\mu)$ is holomorphic everywhere if both $\lambda$ and $\mu$
correspond to the points on the same sheet, but it develops
the second order pole at $\mu=\lambda$, where two points are located on different
sheets. Indeed, taking into account
the only non-holomorphic part gives
\be
W_0(\lambda,\mu)\stackreb{\lambda\to\mu}{\sim}-\sum_k{k\lambda^{k-1}
\over\mu^{k+1}}=-{1\over (\lambda-\mu)^2}
\ee
Besides, the evident normalizing condition, fixing the holomorphic part,
immediately follows from (\ref{BPQ}),
\be\label{a-per}
\oint_{A_i}W_0(\lambda,\mu)d\mu=
\oint_{A_i}W_0(\lambda,\mu)d\lambda=0
\ee
due to \rf{dVSi} and since $W_0(\lambda,\mu)$ is
symmetric in $\lambda$ and $\mu$ by definition.

Therefore, we finally come
to the formula for the 2-point resolvent,
\be
W_0(\lambda ,\mu )d\lambda d\mu =\frac{\partial W_0(\lambda )}
{\partial V(\mu )}d\lambda d\mu =-B(P,Q^*),
\label{*two-loop*}
\ee
where we have introduced the $*$-involution between the two sheets
of the hyperelliptic curve ${\cal C}$, so that $Q^*$ denotes
the image of $Q$ under this involution.
The only singularity of (\ref{*two-loop*}), for a fixed point $P$ on a physical
sheet, is at the point $Q\to P^*$
on the unphysical sheet with $\mu(Q)=\lambda(P^*)=\lambda(P)$, while on the other
sheet it is cancelled under change of sign of $\tilde y$.

Therefore, in order to calculate the 2-point resolvent,
one needs to write down
the Bergmann bidifferential on the hyperelliptic curve manifestly. In principle,
it has
several different representations (one of the most hard for any further
treatment is given by formula
(5.20) in \cite{RG}, borrowed from \cite{LeMo}).
The simplest one can be obtained using formula
(\ref{SzegoB}). Indeed,  a simple
hyperelliptic representation of the Szeg\"o kernel
exists for the even non-singular half-integer
characteristics. Such characteristics are in one-to-one correspondence with
the partitions of the set of all the $2g+2$ ramification points into two
equal subsets, $\{\mu^+_\alpha\}$ and $\{\mu^-_\alpha\}$, $\alpha =
1,\ldots,g+1$, $y_{\pm}(\lambda) \equiv \prod_{\alpha =1}^{g+1} (\lambda -
\mu^{\pm}_\alpha)$, i.e. $y(\lambda)=y_+(\lambda)y_-(\lambda)$.
Given these two sets, one can define $U_e(\lambda) = {y_+(\lambda)\over
y_-(\lambda)}$.
In terms of these functions, the Szeg\"o kernel is
equal to \cite{Fay,MKM}
\be \Psi_e(\lambda,\mu) = \frac{U_e(\lambda) +
U_e(\mu)}{2\sqrt{U_e(\lambda)U_e(\mu)}} \frac{\sqrt{d\lambda d\mu}}{\lambda -
\mu}
\label{Sz}
\ee
Square of this expression, due to \rf{SzegoB}, leads to the manifest expression
for the singular part of 2-point resolvent or the Bergmann kernel, i.e.
\be\label{Berg1}
W_0(\lambda,\mu)d\lambda d\mu=
{y_+^2(\lambda)y_-^2(\mu)+y_+^2(\mu)y_-^2(\lambda)-2y(\lambda)y(\mu)\over
4y(\lambda)y(\mu)}{d\lambda d\mu\over (\lambda-\mu)^2} + \hbox{holomorphic
part}
\ee
where we choose the sign in front of $2y(\lambda)y(\mu)$ in the numerator
so that the pole lies on the unphysical sheet. The holomorphic part is fixed
now by the condition of zero $A$-periods.

However, in our further calculations we need another, completely different
expression for the Bergmann kernel \cite{Ak96}. It can be most
immediately obtained from the loop equation. The loop equation for the
2-point resolvent has the form (see, e.g.,
formula (I.3.40) in \cite{AMM})
\be
V'(\lambda)W_0(\lambda,\mu)-\hat
r_V(\lambda)W_0(\mu)=2W_0(\lambda)W_0(\lambda,\mu)+{\partial\over\partial
\mu}{W_0(\lambda)-W_0(\mu)\over\lambda-\mu}
\ee
where the operator $\hat r_V(\lambda)$ is defined in \rf{R}, i.e. on a hyperelliptic
curve $y^2=R(x)$, e.g. \rf{1mamocu}, one gets
\be\label{Berg3}
W_0(\lambda,\mu)={1\over y(\lambda)}\left[{\partial\over\partial
\mu}{W_0(\lambda)-W_0(\mu)\over\lambda-\mu}+\hat
r_V(\lambda)W_0(\mu)\right]=
\ee
$$
=-{1\over 2(\lambda-\mu)^2}+
{y(\lambda)\over 2 y(\mu)}
\left[{1\over(\lambda-\mu)^2}-{1\over 2 (\lambda-\mu)}
{\partial \log R(\lambda)\over\partial\lambda}-{1\over 4}\hat r_V(\mu)
\log R(\lambda)\right]
$$
One can check by straightforward calculation that this formula leads to a
symmetric expression (see
(III.2.6) in \cite{AMM})
\be\label{Berg2}
W_0(\lambda,\mu)d\lambda d\mu={V'(\mu)V'(\lambda)+{1\over 2}\left(\lambda Q(\mu)+\mu
Q(\lambda)\right) +c-y(\mu)y(\lambda)\over
2y(\lambda)y(\mu)}{d\lambda d\mu\over (\lambda-\mu)^2} + \hbox{holomorphic
part}
\ee
which is a particular case of \rf{Berg1}, when parameterizing the hyperelliptic curve as
\be
y^2(\lambda)=V'^2(\lambda)+\lambda Q(\lambda)+c
\ee
In the case of degenerate Riemann surface \rf{ty}, one obtains (see
\cite{Ak96}) similarly to (\ref{Berg3})
\be
W_0(\lambda,\mu)d\lambda d\mu=-{d\lambda d\mu\over 2(\lambda-\mu)^2}+
\frac{\tilde y(\lambda )}{2\tilde y(\mu)}\left(\frac{1}{(\mu-\lambda )^2}+
\frac12\sum_{\alpha=1}^{2n}\left[\frac{1}{(\mu-\lambda )(\lambda -\mu_\alpha)}
+\frac{{\cal L}_\alpha(\mu)}{\lambda -\mu_\alpha}\right]\right)d\lambda d\mu,
\label{*y(p,q)*}
\ee
where ${\cal L}_\alpha(\mu )\equiv \sum_{l=0}^{n-2}{\cal L}_{\alpha,l} \mu ^l$ are
polynomials in $\mu $. They can be unambiguously fixed by the
requirements of absence of the first-order poles at $\lambda=\mu$ and zero
$A$-periods.

\subsection{Calculating ${\cal L}_\alpha(\mu )$}

Although one has now the explicit formula for the 2-point resolvent
(=Bergmann bidifferential) on the
hyperelliptic surface in terms of branching points (\ref{*y(p,q)*}),
it contains the polynomials ${\cal L}_\alpha(\mu )$ defined
by the implicit requirements.

To find effective formulas for these polynomials,
let us set $\lambda =\mu_\alpha$ in (\ref{*y(p,q)*}),
and introduce the notation
\be
{\tilde y}_\alpha(\lambda )\equiv \sqrt{\prod_{\beta\ne\alpha}(\lambda -\mu_\beta)},
\ \ \ \ \
{\tilde y}_\alpha\equiv \sqrt{\prod_{\beta\ne\alpha}(\mu_\alpha -\mu_\beta)}
\label{*y-alpha*}
\ee
We will also denote by square brackets the {\em fixed} argument
of the Bergmann bi-differential (in some local coordinate), and
consider $B(P,[\mu])$ as a 1-differential on ${\cal C}$.
From (\ref{*two-loop*}), one has
\be
-2B(P,[\mu_\alpha])=\frac12\frac{{\cal L}_\alpha(\lambda )}{{\tilde y(\lambda)}}
{\tilde y}_\alpha^2 d\lambda +
\frac12\frac{{\tilde y}^2_\alpha}{(\lambda -\mu_\alpha){\tilde y(\lambda)}}
d\lambda
\ee
From (\ref{*vanish*}), we have
\be
\oint_{A_i}\frac{{\cal L}_\alpha(\mu )}{{\tilde y}(\mu )}d\mu =-
\oint_{A_i}\frac{d\mu }{(\mu -\mu_\alpha){\tilde y}(\mu )}.
\label{LQ*}
\ee
and integrand in the l.h.s. is a linear combination of canonical holomorphic
differentials $d\omega_i$ (\ref{canoDV}), so that
\be
{\cal L}_\alpha(\mu ) =-\sum_{i=1}^{n-1}
{ H}_i(\mu )\cdot
\oint_{A_i}\frac{d\lambda}{(\lambda-\mu_\alpha){\tilde y(\lambda)}},
\label{*LQ*}
\ee
and, in particular,
\be
{\cal L}_\alpha(\mu_\alpha)=-\sum_{i=1}^{n-1}{ H}_i(\mu_\alpha)\cdot
\oint_{A_i}\frac{d\lambda}{(\lambda-\mu_\alpha){\tilde y(\lambda)}}
\label{*Lmu*}
\ee
Another equivalent representation we will need in sect.~\ref{s:genus}, is
\be
\label{main}
\sum_{l=0}^{n-2}{\cal L}_{\alpha,l}\mu_\alpha^l=-\sum_{j=1}^{n-1}
\oint_{A_j}\frac{ H_j(\lm)}{(\lm-\mu_\alpha)\ty (\lambda)}d\lm,
\quad \alpha=1,\dots,2n.
\ee
Indeed, let us introduce the quantities
\be
\label{Q}
\sigma_{j,i}\equiv\oint_{A_j}\frac{\lm^{i-1}}{\ty
(\lambda)}d\lm,\quad i,j=1,\dots,n-1
\ee
Then, for the
canonical polynomials
$ H_k(\lm)\equiv \sum_{l=1}^{n-1} H_{l,k}\lm^{l-1}$, \ $k=1,\dots,n-1$,
related to the canonically normalized differentials (\ref{canoDV}), i.e.
$\oint_{A_j}\frac{ H_k(\lm)}{\ty (\lambda)}d\lm=\delta_{k,j}$,
one obviously has
\be
\label{inverse}
\sum_{l=1}^{n-1} \sigma_{j,l} H_{l,k}=\delta_{j,k}\quad\hbox{for}\quad j,k=1,\dots,n-1.
\ee
Therefore, for all $k>0$ such that $j-1-k\ge0$,
\bea
&&\sum_{i=1}^{n-1} { H}_{j,i}\cdot
\oint_{A_i}\frac{\lambda^{j-k-1}}{{\tilde y (\lambda)}}d\lambda=0
\eea
Then,
\bea
&&\sum_{i=1}^{n-1} \oint_{A_i}\frac{{ H}_i(\lambda)-{ H}_i(\mu_\alpha)}
{(\lambda-\mu_\alpha){\tilde y (\lambda)}}d\lambda=
\nonumber
\\
&&\quad=
\sum_{i=1}^{n-1} \oint_{A_i}\frac{\sum_{j=2}^{n-1}{ H}_{j,i}
\sum_{k=1}^{j-1}\lambda^{j-1-k}\mu_\alpha^{k-1}}
{{\tilde y (\lambda)}}d\lambda=0.
\nonumber
\eea
and, because of (\ref{*Lmu*}), we finally arrive at (\ref{main}). The above formulas
mean that for the Bergmann kernel on hyperelliptic curve $y^2=R(x)$ one can write
\be
\label{bekehy}
W_0(\lambda,\mu)d\lambda d\mu=
$$
$$
-{d\lambda d\mu\over 2(\lambda-\mu)^2}+
\frac{y(\lambda )}{2y(\mu)}\left(\frac{1}{(\mu-\lambda )^2}+
\frac12\sum_{\alpha=1}^{2n}\left[\frac{1}{(\mu-\lambda )(\lambda -\mu_\alpha)}
-\sum_{i=1}^{n-1}{ H}_i(\mu)
\oint_{A_i}\frac{dx}{(x-\mu_\alpha)^2{y (x)}}\right]\right)d\lambda d\mu =
$$
$$
=-{d\lambda d\mu\over 2(\lambda -\mu)^2}\left(1-{R(\lambda)\over
y(\lambda )y(\mu)}\right) - {d\lambda d\mu\over 2(\lambda -\mu)}
{R'(\lambda)\over y(\lambda )y(\mu)}
- {1\over 2}y(\lambda)d\lambda\sum_{\alpha=1}^{2n}\sum_{i=1}^{n-1}d\omega_i(\mu)
\oint_{A_i}\frac{dx}{(x-\mu_\alpha)^2{y (x)}}
\ee
A particular case of this formula at $\lambda=0$ was used in \cite{KMMZ} for solving
the quasiclassical Bethe anzatz equations in the context of AdS/CFT correspondence.
The last term in the r.h.s. of \rf{bekehy} is explicit form for the action of
the operator \rf{R} in the case of smooth Riemann surface, restricted by the
vanishing period's constraints.

\subsection{Mixed second derivatives
}

Another set of relations follows from the general properties of the Bergmann kernel, and
can be also derived directly from the formulas of sect.~\ref{s:1MM}. To this end,
we apply the mixed derivatives
$\partial /\partial V(\mu )$ and $\d/\d S_i$ to the planar limit of the free energy ${\cal F}_0$. On
one hand, $\d{\cal F}_0/\d S_i=\oint_{B_i}dS$ and using that
$dS(\lambda)=y(\lambda)d\lambda=(V'(\lambda)-2W_0(\lambda))d\lambda$,
${\partial V'(\lambda)\over\partial V(\mu)}=-{1\over (\lambda-\mu)^2}$ and
formula (\ref{*two-loop*}),
one obtains that
\bea\label{abc}
\oint_{B_i}\frac{\partial (dS(\lambda ))}{\partial V(\mu )}&=&\oint_{B_i}
\left(2B(P,[\mu ])-
\frac1{(\lambda -\mu )^2}dp\right)=
\nonumber
\\
&=&
\oint_{B_i}2B(P,[\mu ]).
\nonumber
\eea
On the other hand, acting by derivatives in
the opposite order, one first obtains $\partial {\cal F}_0/\partial V(\mu )=W_0(\mu )=V'(\mu )-y(\mu )$
and then $\d (V'(\mu )-y(\mu ))/\d S_i =2d\omega_i([\mu ])$, or, in the coordinate-free notation,
one of the Fay identities~\cite{Fay}:
\be
{1\over 2\pi i}\oint_{B_i}B(P,Q)=d\omega_i(Q)
\label{*B-Bcycle*}
\ee
This means that
\be\label{dsdt}
{\partial dS(\mu) \over \d S_i}=\left[\oint_{B_i}{\d dS(\lambda )\over\d V(\mu)}d\lambda\right]d\mu=
\oint_{B_i}{\d dS(\mu)\over\d V(\lambda )}d\lambda
\ee
where the {\it both} integrals are taken over the variable $\lambda $.
Now, as $dS(\mu)=y(\mu)d\lambda$ is the generating function for the variables
$\xi_a\equiv
\{{\cal M},\{\lambda_i\},\{\mu_\alpha\},M_\alpha^{(i)}\}$, given by (\ref{ty}), (\ref{M}),
and (\ref{M2}) ($M_\alpha^{(i)}$ are just $i$th order derivatives of $M_{m-n}(\mu )$
at $\mu =\mu_\alpha$) giving their dependence
on $S_i$ and $t_k$, one concludes that similar relation for the first
derivatives holds for each of these variables.
Indeed, multiplying (\ref{dsdt}) by ${1\over
y(\mu)}$ one then can bring $\mu$ successively to $\mu_\alpha$'s, $\lambda_i$'s
and $\infty$ to pick up pole terms with different $\xi_a$ and prove that
\be
{\d \xi_a\over\d S_i}=\oint_{B_i} {\d \xi_a\over\d V(\lambda )}d\lambda
\ee
As a consequence, any function $G$ of $\xi_a$ would naively satisfy the same
relation.
However, there is a subtle point here: in
formula (\ref{*B-Bcycle*}) one could integrate the both sides
over an $A_j$-cycle to obtain at the r.h.s. $\delta_{ij}$, while
the $A$-periods of the Bergmann kernel are zero. It means that one
should carefully permute the integrations, since
$A_i-$ and $B_i$-cycles intersect. Therefore, one should carefully
take into account the contributions of the intersection
points ($A_i$- and $B_j$-cycles for different $i$ and $j$ {\it do not}
intersect and integrations can be exchanged, which perfectly match the
Kronecker symbol obtained) which does not vanish due to the double pole of
the Bergmann kernel.

In particular, we formally have
\be
\label{*mismatch*}
\frac{\partial M(\lambda )}{\partial S_i}=\oint_{B_i}d\mu
\left(\frac{\partial M(\lambda )}{\partial V(\mu )}+\frac{\partial}{\partial \mu }\,\frac{1}{(\lambda -\mu )\ty(\mu )}\right),
\ee
where the second term in the brackets vanishes unless we have an outer integration over
the cycle $A_i$ w.r.t. the variable $\lambda $ in the both sides.

Similarly, one should take care when omitting the second term in the second
equality in (\ref{abc}). Indeed, suppose we consider $S_i$ as a function of
$\xi_a$. Then,
\be
{\partial S_i(\xi_a)\over\d S_j}\ne \oint_{B_i}{\d S_i\over\d V(\lambda )}
\ee
Here again one should note that $S_i$ is the integral of $dS$ over the $A_i$
cycle and take care when exchanging integrations. While doing this, the
contributions from the omitted term in (\ref{abc}) are to be taken into
account. Note, however, that one can {\em never} express $S_i$ as a function
of a {\em finite} number of ``local'' variables $\xi_a$.

Therefore, if one considers the functions $G$ that depends
on only finite number of ``local'' variables $\xi_a$
(the branching points, $M_\alpha^{(i)}$---the
moments of the model, and, possibly, zeros of the polynomial $M(x)$), the relation
\be\label{G}
{\d G\over\d S_i}=\oint_{B_i} {\d G\over\d V(\lambda )}d\lambda
\ee
holds, while, for $A$-cycle integrals over the Riemann surface (note that the
$B$-cycle integrals do not meet such a problem), one should add more
terms in the r.h.s. in order to take into account the second, pole
term in (\ref{abc}), (\ref{*mismatch*}).

Relation (\ref{G}) is therefore valid in {\em all} orders of $1/N$-expansion for the
1MM free energy because any higher-genus contribution is a function only of $\mu_\alpha$
and of a {\em finite} number of higher moments $M_\alpha^{(i)}$.


Equation (\ref{*y(p,q)*}) implies, for $\lambda \sim \mu_\alpha\sim \mu $,
in local coordinates $d\lambda/\sqrt{\lambda -\mu_\alpha}=2d\left(\sqrt{\lambda -\mu_\alpha}\right)$ and
$d\mu/\sqrt{\mu -\mu_\alpha}=2d\left(\sqrt{\mu -\mu_\alpha}\right)$ the relation
\be
{\cal L}_\alpha(\mu_\alpha)=B([\mu_\alpha],[\mu_\alpha])|_{\mathop{nonsing.}}=
S_B(\mu_\alpha).
\ee
Applying now (\ref{*B-Bcycle*}) and (\ref{main}), one immediately
comes to
\be
{\cal L}_\alpha(\mu_\alpha)=
S_B(\mu_\alpha)=\sum_{i=1}^{n-1}\oint_{A_i}\oint_{B_i}\frac{B(P,Q)}{\lambda -\mu_\alpha}
\label{*SB*}
\ee
in the local coordinates associated with the hyperelliptic
Riemann surface (\ref{ty}).

\section{WDVV equations \label{ss:WDVV}}

The
general form
of the Witten--Dijkgraaf--Verlinde--Verlinde (WDVV)
equations~\cite{Wit90,LGMT}
is the systems of algebraic equations~\cite{wdvvg}
\be
\label{WDVV}
\F_I{\F}_J^{-1}\F_K = \F_K{\F}_J^{-1}\F_I, \quad \forall\ I,J,K
\ee
on the third derivatives
\be
\label{matrF}
\|{\F}_{I}\|_{JK}=
{\d^3\F\over\d t_I\,\d t_J\,\d t_K} \equiv\F_{IJK}
\ee
of some function $\F (\{t_I\})$. These equations often admit
an interpretation as associativity relations in some algebra (of
polynomials, differentials etc.)
and are relevant for describing topological theories.

As WDVV systems are often closely related to Whitham systems,
a natural question is whether the corresponding 1MM free-energy function
satisfy the WDVV equations?
This was proved in \cite{ChMMV}, where it was shown that
the multicut solution the 1MM satisfies the WDVV
equations as a function of {\em canonical} variables identified with
the periods and residues of the generating meromorphic one-form $dS$
\cite{Kri1}. The method to prove it consists of two steps.
The first, most difficult, step is to find the residue formula for the third
derivatives (\ref{matrF}) of the 1MM free energy. Then, using an
associativity, one immediately proves that the
free energy of multi-support solution satisfies
the WDVV equations if the number of independent variables
is {\it equal\/} to the number of branching (critical) points in the residue formula.
We show here that the statement holds in the case of arbitrary potentials
for a fixed-genus reduced Riemann surface.

In sect.~\ref{ss:residue}, we derive the residue formula
for the third derivatives of the quasiclassical tau-function for the
variables (the generalized times) $t_I$ associated
with both the periods $S_i$ and residues
$t_k$ of the generating differential $dS$.
In sect.~\ref{ss:proof}, we prove
that the free energy of the multi-interval-support solution ${\cal F}_0 (t_I)$
solves WDVV equations (\ref{WDVV}) as a function of the subset
$\{t_\alpha\}\subseteq \{t_I\}$; the total number of $t_\alpha$ must
be fixed to be equal to the number of branching points in the residue
formula for the third derivatives (\ref{matrF}) in order to
make the set of the WDVV equations nontrivial.

\subsection{Residue formula
\label{ss:residue}}


Because all the quantities $d\Omega_I$ (\ref{dOI}) depend {\it entirely\/}
on the reduced hyperelliptic Riemann surface (\ref{ty}), their
derivatives w.r.t. {\it any parameter\/} must be expressed through
the derivatives w.r.t. the positions of the branching points $\mu_\alpha$.
So, calculating derivatives w.r.t.\ $\mu_\alpha$
is the basic ingredient. Note that
although the differentials $d\Omega_I$ are regular at the points
$\mu_\alpha$ in the local coordinate $d\tilde y\sim
d\lambda/\sqrt{\lambda-\mu_\alpha}$,
the derivatives $\d d\Omega_I/\d
\mu_\alpha$ obviously develop singularities at $\lambda=\mu_\alpha$,
and we must bypass these singularities when choosing the integration
contour as in Fig.~\ref{fi:cut}.

Let us now derive the formulas for the third derivatives
$\d^3\F_0/(\d t_I\d t_J\d \mu_\alpha)\equiv \F_{IJ\alpha}$, following
\cite{Kri1,ChMMV}.
Consider, first, the case where the ``times'' $t_I$ and $t_J$ are
$S_i$ and $S_j$, and $\frac{\d^2 \F_0}{\d t_I\d t_J}\equiv\F_{IJ}=T_{ij}$.
We note that the derivatives of the elements
of period matrix
can be expressed through the integral over the ``boundary"
$\d\Sigma_g$ of the cut Riemann surface $\Sigma_g$ (see Fig.~\ref{fi:cut}),
Indeed,
because of the normalization condition $\oint_{A_l}d\omega_j=\delta_{ij}$,
we have $\oint_{A_l}\d_\alpha d\omega_j=0$, so that
\bea
{\d T_{ij}\over \d \mu_\alpha} &=& \oint_{B_j}\d_\alpha
d\omega_i =
\sum_{l=1}^g\left(\oint_{A_l}d\omega_j\oint_{B_l}\d_\alpha
d\omega_i  - \oint_{B_l}d\omega_j\oint_{A_l}\d_\alpha d\omega_i\right)=
\nonumber
\\
&=&
\sum_{l=1}^g\left(\int_{B_l}\omega_j^+\d_{\alpha}d\omega_i -
\int_{B_l}\omega_j^-\d_{\alpha}d\omega_i\right) -
\nonumber
\\
&&\quad\quad-\sum_{l=1}^g\left(\int_{A_l}\omega_j^+\d_{\alpha}d\omega_i -
\int_{A_l}\omega_j^-\d_{\alpha}d\omega_i\right) =
\nonumber
\\
&=&\oint_{\d\Sigma_g}\omega_j\d_{\alpha} d\omega_i
\label{dpm}
\eea
where
$\omega_j = \int d\omega_j$ are Abelian integrals and we let
$\omega_j^\pm$ denote their values on two copies
of cycles on the cut Riemann surface in Fig.~\ref{fi:cut}.

We must now choose the cycles
$A_l$ and $B_l$ bypassing all possible
singularities of the integrand (in this case, the ramification
points $\mu_\beta$).
Expression (\ref{dpm}) can be then evaluated through the residue
formula
\be
\label{dpmres}
\d_\alpha T_{ij} = -\int_{\d\Sigma_g}\omega_j\d_{\alpha} d\omega_i =
\sum\res_{d\lambda = 0} \left(\d_\alpha \omega_j d\omega_i\right).
\ee
The proof of this formula for generic variation of moduli can be found
in \cite{ChMMV}. Here we will adjust it for the class of variations in
terms of the branch points $\{\mu_\alpha\}$, i.e. to the class of
hyperelliptic Riemann surfaces.

Before evaluating this sum of residues, let us consider
the case of meromorphic differentials. Then, using formulas (\ref{v41}) and
(\ref{dfdt0}), one obtains
\be
{\d^2\F_0 \over\d t_k\d t_l} =
\frac12\oint_{C_L} \left((\Omega_k)_{+,0} d\Omega_l\right),
\quad k,l\ge0,
\ee
where
$(\Omega_k)_{+,0}$ is the singular part of $\Omega_k$ at infinity, i.e.,
it is $\lambda^k$ for $k>0$ and the logarithmic function
for $k=0$. Because $\d (\Omega_k)_{+,0}/\d \mu_\alpha=0$ for $k\ge0$,
we have $\d d\Omega_k/\d \mu_\alpha=
\d (d\Omega_k)_-/\d \mu_\alpha$, where $(d\Omega_k)_-$ is the
holomorphic part of $d\Omega_k$ at infinity
(and the expression $(\Omega_k)_-$ is therefore meromorphic for all
$k\ge0$). We then have
\bea
&&{\d\over\d \mu_\alpha}
\frac12\oint_{C_L} \left((\Omega_k)_{+,0} d\Omega_l\right) =
\frac12\oint_{C_L} \left((\Omega_k)_{+,0} \d_\alpha(d\Omega_l)_-\right) =
\nonumber
\\
&&\quad\quad=-\frac12\oint_{C_L} \left(d\Omega_k \d_\alpha(\Omega_l)\right) =
-\res_\infty \left(d\Omega_k \d_\alpha \Omega_l\right).
\eea
The last expression can be rewritten as
\bea
-\res_\infty \left(d\Omega_k {\d \Omega_l\over \d \mu_{\alpha}}\right) &=&
\oint_{\d\Sigma_g}\left(d\Omega_k {\d \Omega_l\over \d \mu_{\alpha}}\right) +
\sum_{\beta=1}^{2n}
\res_{\mu_\beta} \left(d\Omega_k \d_\alpha \Omega_l\right)=
\nonumber
\\
&=&\sum_{\beta=1}^{2n}
\res_{\mu_\beta} \left(d\Omega_k \d_\alpha \Omega_l\right)
\label{3mero}
\eea
because $\oint_{\d\Sigma_g}\left(d\Omega_k
\d_\alpha \Omega_l\right) =0$ following the same arguments
as in formula (\ref{dpm}) and due to normalization
conditions (\ref{v31}).

Further computations for both holomorphic and meromorphic differentials
coincide as we need only their local behavior at the vicinity of a
point $\lambda=\mu_\beta$. Using explicit expression
(\ref{dOI}), we have
\be
\label{dOmega}
\d_\alpha\Omega_J={ H_{J}(\mu_\alpha)\over
\prod_{\gamma\ne\alpha}\sqrt{\mu_\alpha-\mu_\gamma}}
(\lambda-\mu_\alpha)^{-1/2}+O(\sqrt{\lambda-\mu_\alpha}),
\ee
for $\beta=\alpha$, and $\d_\alpha\Omega_J\sim
\sqrt{\lambda-\mu_\beta}$ otherwise. Together with (\ref{dOI}), this
means that the only point to evaluate the residue is
$\lambda=\mu_\alpha$ at which we have (cf.~\cite{Fay})
\be
\label{Fay}
\F_{0,IJ\alpha}=\res_{\mu_\alpha}\bigl(d\Omega_I\d_\alpha\Omega_J\bigr)
={ H_{I}(\mu_\alpha)
 H_{J}(\mu_\alpha)\over \prod_{\beta\ne\alpha}
(\mu_\alpha-\mu_\beta)}.
\ee

Completing the calculation
of the third derivative needs just inverting the dependence on the ramification
points therefore finding $\d \mu_\alpha/\d t_K$. Differentiating
expressions (\ref{1mamocu}), (\ref{ty}) w.r.t. $t_K$ for computing (\ref{ds}) we obtain
\bea
&&{\d dS\over \d t_K}={ H_{K}(\lambda)d\lambda\over \tilde y(\lambda)}
=
\nonumber
\\
&&\quad\quad=\frac12 M_{m-n}(\lambda)\sum_{\alpha=1}^{2n}
{{\tilde y}(\lambda)\over
(\lambda-\mu_\alpha)}{\d \mu_\alpha\over \d t_K}d\lambda+
{\d M_{m-n}(\lambda)\over\d t_K}\tilde y(\lambda)d\lambda.
\label{diff1}
\eea
The derivative of the polynomial $M_{m-n}(\lambda)$ is obviously polynomial
and regular at $\lambda=\mu_\alpha$. Multiplying (\ref{diff1})
by $\sqrt{\lambda-\mu_\alpha}$ and setting $\lambda=\mu_\alpha$,
we immediately obtain
\be
\label{diff2}
\frac{\d \mu_\alpha}{\d t_K}=\frac{ H_{K}(\mu_\alpha)}
{M_{m-n}(\mu_\alpha)\prod_{\beta\ne\alpha}
(\mu_\alpha-\mu_\beta)}.
\ee
Combining this with (\ref{Fay}), we come to the desired residue formula for the
third derivative w.r.t. the canonical variables $t_I$:
\be
\label{resgen}
{\d^3 \F_0\over \d t_I\d t_J\d t_K} =
\sum_{\alpha=1}^{2n}{ H_I(\mu_{\alpha})
 H_J(\mu_{\alpha}) H_K(\mu_{\alpha})\over
M_{m-n}(\mu_\alpha)
\prod_{\beta\ne\alpha}(\mu_{\alpha}-\mu_{\beta})^2}=
\sum_{\alpha}\res_{\mu_{\alpha}}{d\Omega_Id\Omega_Jd\Omega_K\over d\lambda
dy}=\res_{d\lambda=0}{d\Omega_Id\Omega_Jd\Omega_K\over d\lambda dy}
\ee

\subsection{Proof of WDVV equations\label{ss:proof}}

Given residue formula (\ref{resgen}), the proof of WDVV equations
(\ref{WDVV}) can be done, following \cite{wdvvg}-\cite{wdvvmore},
by checking associativity of the algebra
of differentials $d\Omega_I$ with multiplication modulo $\tilde
y(\lambda)d\lambda$. This algebra is reduced to the algebra of polynomials
$ H_I(\lambda)$ with multiplication modulo $\tilde
y^2(\lambda)$ which is correctly defined and
associative. The basis of the algebra of $ H_I(\lambda)$
obviously has dimension $2n$ and is given, e.g., by monomials of
the corresponding degrees $0,1,\dots,2n-2,2n-1$. The study of such algebras
can be performed even for non-hyperelliptic curves, the details can be found
in \cite{wdvvlong,Luuk}.

Another proof is even more simple and reduces to solving the system of linear equations
\cite{BMRWZ,MaWDVV}. To this end, we first define
\be\label{phi}
\phi_I^{\alpha}\equiv
{ H_I(\mu_{\alpha})
\over
M^{1/3}_{m-n}(\mu_\alpha)
\prod_{\beta\ne\alpha}(\mu_{\alpha}-\mu_{\beta})^{2/3}}
\ee
so that (\ref{resgen}) can be rewritten as
\be\label{Fphi3}
\F_{0,IJK}=\sum_{\alpha} \phi_I^{\alpha}\phi_J^{\alpha}\phi_K^{\alpha}
\ee

Now let us fix some index $Y$ and consider the following multiplication
\be
\label{eqc}
\phi_I^{\alpha}\phi_J^{\alpha} =\sum_K
C^{(Y)K}_{IJ}\phi_K^{\alpha}\phi_Y^{\alpha}, \quad \forall\ \alpha
\ee
the structure constants $C^{(Y)K}_{IJ}$ being independent of
$\alpha$.
One can equally look at this as at
a system of {\em linear equations} for $C_{IJ}^K$ at fixed values of $I$ and
$J$. If this system has a solution, (\ref{eqc}) gives rise to an associative
ring, with the structure constants $C_{IJ}^K$ satisfying (associativity
condition)
\be\label{CC}
\left(C^{(Y)}_I\right)^K_L \left(C^{(Y)}_J\right)^L_M = \left(C^{(Y)}_J\right)^K_L
\left(C^{(Y)}_I\right)^L_M, \quad
(C^{(Y)}_I)_J^K \equiv C^{(Y)K}_{IJ}
\ee

Now, the solution to (\ref{eqc}) is
\be
\label{litc}
C^{(Y)K}_{IJ} = \sum_\alpha
\phi_I^{\alpha}\phi_J^{\alpha}
\left(\Phi_{(Y)K}^{\alpha}\right)
\ee
where $\Phi_K^{\alpha}$ is the matrix\footnote{We consider it as a matrix of
indices $K$ and $\alpha$, while the reference index $Y$ is implied as a
silent parameter of the whole consideration.
} inverse to
$\phi_K^{\alpha}\phi_Y^{\alpha}$.
This solution exists if the number of vectors (variables $K$) is
{\it greater or equal\/} the number $2n$ of the branching points
$\mu_\alpha$.

The other important condition is the
{\it invertibility\/} of the matrix $\phi_K^{\alpha}\phi_Y^{\alpha}$, or the
matrix $\phi_K^{\alpha}$ (we suppose $\phi_Y^{\alpha}\ne 0$)
which ensures the nontriviality of the WDVV relations.
For this, we must require the number of vectors
$\phi_I$ to be {\it less or equal\/} the number $2n$ of their
components.  We therefore obtain the following two conditions \cite{MaWDVV}:
\begin{itemize}

\item the ``matching" condition
\be
\label{matching}
\#(I)=\#(\alpha);
\ee
and

\item the nondegeneracy of the matrix $\phi_I^{\alpha}$
(see the proof in sect.~\ref{ss:torus}):
\be
\label{det}
\det_{I\alpha}\| \phi_{I}^{\alpha}\| \neq 0
\ee
\end{itemize}

Now, using (\ref{eqc}), one rewrites (\ref{Fphi3})
\be
\F_{0,IJK}=\sum_{\alpha,L} C^{(Y)L}_{IJ}\phi^{\alpha}_K\phi^{\alpha}_L\phi^{\alpha}_Y=
\sum_L C^{(Y)L}_{IJ} \F_{0,KLY}
\ee
coming to the matrix formula that express the structure constants through
the third derivatives of the planar limit free energy
\be\label{C}
C^{(Y)}_{I}=\F_I \F^{-1}_Y
\ee
where we denoted $\F_I$ the matrix with element $JK$ equal to $\F_{0,IJK}$.
Now substituting (\ref{C}) into (\ref{CC}), one immediately arrives at
(\ref{WDVV}).

Thus, we established that conditions (\ref{WDVV}) require
the number of varying parameters $\{ t_I\}$ to satisfy matching
condition (\ref{matching}).
We let $\{t_\alpha\}$ denote these ``primary'' variables.
It is convenient to set classical ``primary'' variables w.r.t.
which WDVV equations (\ref{WDVV}) hold true,
to be the parameters $S_i$, \
$i=1,\dots,n-1\equiv g$, \ $t_0$, and $t_k$ with $k=1,\dots,n$ keeping
all other times frozen.

Below, in sect.~\ref{ss:torus}, we interpret the answer in genus one
in terms of the determinant relation for the third derivatives ${\cal F}_{IJK}$.

\section{Higher genus contributions \label{s:genus}}

The solution $W_1(\lambda )$ to the loop equations in the multicut case
was first found by
Akemann~\cite{Ak96}.\footnote{The universal critical behavior of the
corresponding correlation functions was discussed in~\cite{AkAm}.}
He also managed to integrate them in to obtain the free energy $\F_1$ in the
two-cut case.
The genus-one partition function in the generic multi-cut
case was proposed in \cite{Kos,DST}, where it was observed that the Akemann formula
coincides with the correlator of twist fields, computed by Al.Zamolodchikov
\cite{Zam}. This produces cuts on complex plane and gives rise to a hyperelliptic
Riemann surface, following the ideology of \cite{Knizhnik},
some corrections to this construction are due to
the star operators, introduced in \cite{Moore}.
In this section, we present (see also \cite{Chekh}) the
derivation of the genus-one correction based on solving the loop equation,
and generalizing Akemann's result for the partition function to arbitrary number
of cuts.

\subsection{The iterative procedure}\label{iterate}

\paragraph{Iterative solving the loop equations.}

Thus, now we are going to determine
higher genus contributions. We do this iteratively by inverting
the genus expanded loop equation (\ref{4.11}).
Our strategy will be to
construct an integral operator $\widehat d{\cal G}$ inverse to the integral operator
$\widehat{K}-2W_0(\lambda)$.

Acting with this operator onto the both sides of the loop equation eq.(\ref{4.11}), one
recursively produces $W_h(\lambda )$ for all genera like all the
multi-point resolvents of the same genus can be obtained from $W_h(\lambda )$ merely
applying the loop insertion operator $\dV$.

However, there is a subtlety: the operator $\widehat{K}-2W_0(\lambda)$ has
zero modes and is not invertible. Therefore, solution to the loop equation
is determined up to an arbitrary combination of these zero modes.
Hence, the kernel of the operator $\widehat{K}-2W_0(\lambda)$ is
spanned exactly by holomorphic one-differentials on the Riemann surface (\ref{ty}).

In order to fix this freedom, we assume $W_h(\lambda)$ is expressed exclusively in
terms of derivatives $\frac{\d \mu_\alpha}{\d V(\lambda)}$ and
$\frac{\d M_\alpha^{(k)}}{\d V(\lambda)}$,
which, as we show in the next paragraph, fixes a solution to the loop
equation. It is a natural extension of the normalizing property (\ref{I}) to
higher genera and can be ultimately written in the form
\be
\label{A-cycle-F}
\oint_{A_i}\frac{\d \F_h}{\d V(\lambda)}d\lambda\equiv
\oint_{A_i}W_h(\lambda)d\lambda=0\quad \forall i\hbox{\ and\ for\ }h\ge1.
\ee
Now we claim that the integral operator
\be
\label{1*}
\widehat{d{\cal G}}\left(f\right)(\lambda)\equiv
\oint_{{\cal C}_{\mu_\alpha}}\frac{d\mu }{2\pi i}\,\frac{d{\cal G}(\lambda ,\mu )}{d\lambda }\frac{1}{y(\mu )}
\cdot f(\mu )
\ee
is an inverse for the operator $\widehat{K}-2W_0(\lambda)$ in the space of
rational functions $f(\mu)$ with poles at the points $\mu_{\alpha}$ only.

Here
the one-differential $d{\cal G}(\lambda ,\mu )$ w.r.t. the first argument $\lambda $
\footnote{It is the function $dS(\lambda ,\mu )$ in the notation of \cite{Ey}.}
is the primitive of the Bergmann kernel
$B(\lambda ,\mu )$ w.r.t. the argument $\mu $. Obviously, it is a single-valued differential of $\lambda $ with
zero $A$-periods on the reduced Riemann surface and is multiple-valued function of $\mu $,
which undergoes jumps equal to $d\omega_i(\lambda )$ when the variable $\mu $ passes
through the cycle $B_i$ (cf. with \rf{bekehy}):
\be
\label{*dE}
d{\cal G}(\lambda ,\mu )=\frac{\ty(\mu )d\lambda }{(\lambda -\mu )\ty(\lambda )}-
\sum_{i=1}^{n-1}\frac{ H_{i}(\lambda )d\lambda }{\ty(\lambda )}\oint_{A_i}d\xi
\frac{\ty(\mu )}{(\xi-\mu )\ty(\xi)}.
\ee
Contours of integration over the cycles $A_i$ must lie outside the contour
of integration ${\cal C}_{\mu_\alpha}$ encircling the branching point
$\mu_\alpha$ in (\ref{*1*}).

Moreover, this operator obeys the property
\be\label{normG}
\oint_{A_i}\widehat{d{\cal G}}(f)(\lambda )d\lambda \equiv 0
\ee
and, therefore, respects condition (\ref{A-cycle-F}). Therefore, one has to
solve the loop equations inverting $\widehat{K}-2W_0(\lambda)$ exactly with
$\widehat d{\cal G}$.

Then, the calculation immediately validates the diagrammatic technique \cite{Ey} for
evaluating multipoint resolvents in 1MM. Indeed,
representing $d{\cal G}(\lambda ,\mu )$ (for $\lambda >\mu $) as the arrowed propagator,
the three-point vertex as dot in which
we assume the integration over $\mu $:
$\bullet\equiv\oint\frac{d\mu }{2\pi i}\frac{1}{y(\mu )}$, we can graphically write solution
to (\ref{4.11}) since
\be
\label{diaey}
W_h(\lambda )=\widehat{d{\cal G}}\left[\sum_{h'=1}^{h-1}W_{h'}(\cdot)W_{h-h'}(\cdot)+W_{h-1}
(\cdot,\cdot)\right](\lambda ).
\ee
Then, representing multiresolvent $W_{h'}(\lambda_1,\dots,\lambda_k)$ as the block with
$k$ external legs and with the index $h'$,
one obtains
\be
\begin{picture}(190,55)(10,10)
\thicklines
\put(40,40){\oval(20,20)}
\thinlines
\put(20,40){\line(1,0){12}}
\put(20,40){\circle*{2}}
\put(20,44){\makebox(0,0)[cb]{$\lambda $}}
\put(40,40){\makebox(0,0)[cc]{$h$}}
\put(55,40){\makebox(0,0)[lc]{$=\sum\limits_{h'=1}^{h-1}$}}
\put(80,40){\vector(1,0){13}}
\put(80,40){\circle*{2}}
\put(95,40){\circle*{4}}
\put(80,44){\makebox(0,0)[cb]{$\lambda $}}
\put(93,44){\makebox(0,0)[cb]{$\mu $}}
\put(95,40){\line(1,1){7}}
\put(95,40){\line(1,-1){7}}
\thicklines
\put(107,52){\oval(20,20)}
\put(107,28){\oval(20,20)}
\put(107,52){\makebox(0,0)[cc]{$h-h'$}}
\put(107,28){\makebox(0,0)[cc]{$h'$}}
\thinlines
\put(120,40){\makebox(0,0)[lc]{$+$}}
\put(130,40){\vector(1,0){13}}
\put(130,40){\circle*{2}}
\put(145,40){\circle*{4}}
\put(130,44){\makebox(0,0)[cb]{$\lambda $}}
\put(143,44){\makebox(0,0)[cb]{$\mu $}}
\put(145,40){\line(1,1){9}}
\put(145,40){\line(1,-1){9}}
\thicklines
\put(166,40){\oval(30,30)}
\put(166,40){\makebox(0,0)[cc]{$h-1$}}
\put(185,40){\makebox(0,0)[cc]{$,$}}
\end{picture}
\label{Bertrand}
\ee
which is just the basis relation for the diagrammatic representation \rf{diaey}. Here,
by convention, all integration contours for the variables $\mu $ lie inside each other
in the order, established by arrowed propagators $d{\cal G}(\mu_i,\mu_j)$. The other,
nonarrowed propagators are $W(\lambda ,\mu )\equiv \frac{\d}{\d \mu }
d{\cal G}(\lambda ,\mu )$. All $A$-cycle contours of integration in
$d{\cal G}(\lambda ,\mu )$
are outside the contours of internal integrations over $\mu_i$-variables.

\paragraph{Choosing a specific basis.}

In order to prove the claims of the previous paragraph,
first of all, we change variables from coupling constants
to special moment functions which
allows one to apply higher genus machinery nonperturbatively
in coupling constants $t_j$. This machinery turns out to involve
only on a finite number of the moments $M_\alpha^{(k)}$ (\ref{M1}).
(Recall that (\ref{M})-(\ref{M1}) implies $M_\alpha^{(1)}=M(\mu_\alpha)$.)

Now let us fix the generic analytic structures of the 1- and 2-point
resolvents. First of all, note that $W_0(\lambda,\mu)$ in (\ref{Berg3}) is
invariant w.r.t. the involution $y\to -y$ that permutes physical and
unphysical sheets. Moreover, $W_0(\lambda,\lambda)$ is a
fractional rational function of $\lambda$ with poles at the points
$\mu_{\alpha}$ only. Further, look at the loop equation, (\ref{4.11}) and
rewrite it as
\beq
y(\lambda)W_h(\lambda)=\left[V'(\lambda)W_h(\lambda)\right]_++
\sum_{h'=1}^{h-1}
W_{h'}(\lambda)W_{h-h'}(\lambda)+\frac{\partial }{\partial V(\lambda)}W_{h-1}(\lambda),
\label{4.11*}
\eeq
It follows from this formula that the 1-point resolvent $W_1(\lambda)$ is also
fractional rational function of $\lambda$ with poles at the points
$\mu_{\alpha}$ only divided by $y(\lambda)$, i.e. it changes sign
under permuting the physical and unphysical sheets. This procedure can be
iterated with the loop equation written in the form \cite{AMM}
$$
V'(\lambda)W_h(\lambda,\lambda_1,...,\lambda_n)=
\hat r_V(\lambda) W_h(\lambda_1,...,\lambda_n)+
\sum_{h'=0}^{h}\sum_{n_1+n_2=n-1}
W_{h'}(\lambda,\lambda_1,...,\lambda_{n_1})
W_{h-h'}(\lambda,\lambda_1,...,\lambda_{n_2})+
$$
\beq
+\sum_i
{\partial\over\partial\lambda_i}{W_h(\lambda,\lambda_1,...,\check\lambda_i,...,\lambda_n)
-W_h(\lambda_1,...,\lambda_n)\over\lambda -\lambda_i}+
\frac{\partial }{\partial V(\lambda)}W_{h-1}(\lambda,\lambda_1,...,\lambda_n)
\label{4.11**}
\eeq
In particular, using the known analytic structure of $W_0(\lambda,\mu)$ one
easily checks that the 3-point resolvent $W_0(\lambda,\mu,\nu)$ is odd
w.r.t. to the involution $y\to -y$ and then, with the knowledge of structure
of $W_1(\lambda)$ and $W_0(\lambda,\mu)$, one can use the loop equation
(\ref{4.11**}) to prove that $W_1(\lambda,\mu)$ is even w.r.t. the
involution etc. The final result is that all $(2n)$-point resolvents $W_h$
are even, while all $(2n+1)$-point resolvents
$W_h$ but $W_0(\lambda)$ are odd w.r.t. the involution. This means that
all $W_h(\lambda)$ but $W_0(\lambda)$ are
fractional rational functions of $\lambda_i$ with poles at the points
$\mu_{\alpha}$ only divided by $y(\lambda)$. Moreover,
the r.h.s. of eq. (\ref{4.11}) is similarly a fractional rational
function of $\lambda$ having poles at $\mu_\alpha$ only,
i.e. one should naturally choose a specific basis
$\chi_\alpha^{(k)}(\lambda)$ defined by the property that,
for the integral operator in eq.(\ref{4.11}),
\bea
&&(\widehat{K}-2W_0(\lambda)) \chi_\alpha^{(k)}(\lambda) =\frac{1}{(\lambda-\mu_\alpha)^k},
\nonumber
\\
&&\quad k=1,2,\dots ,\quad \alpha=1,\ldots,2n\, .
\label{Basis}
\eea
Then, $W_h(\lambda)$ must
have the structure
\be
W_h(\lambda)  = \sum_{k=1}^{3h-1}\sum_{\alpha=1}^{2n} A_{\alpha,h}^{(k)} \chi_\alpha^{(k)}(\lambda),
\quad h\ge 1,
\label{Wstr}
\ee
where $A_{\alpha,h}^{(k)}$ are certain functions of $\mu_\beta$ and the
moments $M_\beta^{(k)}$. As the order of the highest singularity term
$1/\bigl((\lambda-\mu_\alpha)^{3h-1}\ty(\lambda)\bigr)$ in $W_h(\lambda)$
is insensitive to a multi-cut structure\footnote{This can be also stated
from the analysis of the loop equations as above.
}, $W_h(\lambda)$ will
depend on at most $2n(3h-2)$ moments, just like the one-cut solution case \cite{ACKM}.

One could define a set of basis functions $\chi_\alpha^{(k)}(\lambda)$ recurrently, as in
\cite{ACKM}, \cite{Ak96}, however,
here we present another technique inspired by \cite{Ey}. We first calculate the quantities
$\frac{\d \mu_\alpha}{\d V(\lambda)}$ and $\frac{\d M_\alpha^{(k)}}{\d V(\lambda)}$.

Using the identity $\parV \Vp(\mu)  = -\frac{1}{(\lambda-\mu)^2}$ and
representation (\ref{M1}),
one easily obtains
\bea
\frac{\d M_\alpha^{(k)}}{\d V(\lambda)}&=&
     (k+{1}/{2}) \left( M_\alpha^{(k+1)}
\dmul-\frac{1}{(\lambda-\mu_{\alpha})^{k+1}\tilde y(\lambda)}\right)\nonumber\\
  &&+\ \frac{1}{2} \sum_{\beta=1 \atop \beta\not= \alpha}^{2n}
      \sum_{l=1}^k \frac{1}{(\mu_\beta-\mu_\alpha)^{k-l+1}} \Big(
\frac{1}{(\lambda-\mu_{\alpha})^{l}\tilde y(\lambda)}-
          M_\alpha^{(l)} \frac{\d\mu_\beta}{\d V(\lambda)} \Big)                  \nonumber\\
  &&+\ \frac{1}{2} \sum_{\beta=1 \atop \beta\not= \alpha}^{2n}
        \frac{1}{(\mu_\beta-\mu_\alpha)^k}
           \Big( M_\beta^{(1)}\frac{\d\mu_\beta}{\d V(\lambda)}-\frac{1}{(\lambda-
\mu_{\beta})\tilde y(\lambda)} \Big)\\ \nonumber
&& \ \ \alpha=1,\ldots,2n \ , \ k=1,2,\dots .
\label{dM}
\eea
Note that the general structure of this formula is
\be\label{dM*}
\frac{\d M_\alpha^{(k)}}{\d V(\lambda)}=
\left(...\right)\dmul+\sum_{\beta\neq\alpha}\left(...\right)
   \frac{\d \mu_\beta}{\d V(\lambda)}-
\frac{\d}{\d \lambda}\left(\frac{1}{(\lambda-\mu_\alpha)^k\ty(\lambda)}\right),
\ee

In order to calculate the derivative of the branching point $\mu_{\alpha}$
w.r.t. the potential, one can note that $W_0(\lambda,\mu)={1\over 2}
{\partial (V'(\mu)-y(\mu))\over\partial V(\lambda)}$ and bring the variable
$\mu$ in this expression to $\mu_{\alpha}$. Then, from
(\ref{*y(p,q)*}) and (\ref{*LQ*}), one obtains that
\be
\label{dmuVp}
M_\alpha^{(1)}\frac{\d\mu_\alpha}{\d V(\lambda)}=
\frac{1}{(\lambda-\mu_\alpha)\ty(\lambda)}-\sum_{i=1}^{n-1}\frac{ H_i(\lambda)}{\ty(\lambda)}
\oint_{A_i}\frac{d\xi}{(\xi-\mu_\alpha)\ty(\xi)}.
\ee

It immediately follows from these formulas and (\ref{LQ*}), (\ref{*LQ*})
that integrals over $A$-cycles
of both $\frac{\d \mu_\alpha}{\d V(\lambda)}$ and $\frac{\d M_\alpha^{(k)}}{\d V(\lambda)}$
vanish and one, therefore, arrives at (\ref{A-cycle-F}).

Using the above conditions and formula (\ref{A-cycle-F}), we can now invert the operator
$\widehat K-2W_0(\lambda)$ when acting on basis monomials $(\lambda-\mu_\alpha)^{-k}$.
That is, we are going to check that the basis $\chi_\alpha^{(k)}(\lambda)$ vectors
(\ref{Basis}) are generated from these basis monomials by the operator
$\widehat {d{\cal G}}$
\be
\label{*1*}
\chi_\alpha^{(k)}(\lambda )=\oint_{{\cal C}_{\mu_\alpha}}
\frac{d\mu }{2\pi i}\,{1\over y(\mu)}\frac{d{\cal G}(\lambda ,\mu )}{d\lambda}
\cdot\frac{1}{(\mu -\mu_\alpha)^k}
\equiv \widehat{d{\cal G}}\left((\lambda -\mu_\alpha)^{-k}\right),
\ee

First few basis functions are easy to obtain from (\ref{dmuVp}) and (\ref{*dE}):
\bea
\chi_\alpha^{(1)}(\lambda) &=& \dmul,\quad \alpha=1,\ldots,2n,
\nonumber\\
\chi_\alpha^{(2)}(\lambda) &=& -\frac{2}{3}\dV \log |M_\alpha^{(1)}|-
\nonumber
\\
&&\quad\quad
-\frac{1}{3}\sum_{\beta=1 \atop \beta\neq \alpha}^{2n} \dV \log |\mu_\alpha-\mu_\beta|.
\label{basis2}
\eea

\paragraph{Proof of (\ref{1*}).}

First, let us demonstrate that the action of the operator $\widehat{d{\cal G}}$ defined in (\ref{*1*}) on any basis
function $(\lambda -\mu_\alpha)^{-k}$ inverts the action of $\widehat K-2W_0(\lambda )$ up to the zero mode content. For this,
let us consider the expression
\bea
\nonumber
&&(\widehat K-2W_0(\lambda ))\oint_{{\cal C}_{\cal D}}\frac{d\mu }{2\pi i}\frac{d{\cal G}(\lambda ,\mu )}{y(\mu )(\mu -\mu_\alpha)^k}
\\
&=&\oint_{{\cal C}_{{\cal D}_w}}\frac{dw}{2\pi i}\frac{V'(w)}{\lambda -w}\oint_{{\cal C}_{{\cal D}_\mu }}
\frac{d\mu }{2\pi i}\frac{d{\cal G}(w,\mu )}{dw}\frac{1}{y(\mu )(\mu -\mu_\alpha)^k}-
\nonumber
\\
&&\qquad\qquad-2W_0(\lambda )\oint_{{\cal C}_{{\cal D}_\mu }}\frac{d\mu }{2\pi i}\frac{d{\cal G}(\lambda ,\mu )}{d\lambda }
\frac{1}{y(\mu )(\mu -\mu_\alpha)^k}.
\nonumber
\eea
Taking into account that the contour ordering is such that ${\cal C}_{{\cal D}_w}>{\cal C}_{{\cal D}_\mu }$
in the sense that one lies inside the other and evaluating the integral over $w$ by taking residues at the
points $w=\lambda $ and $w=\infty$, we find that the result of the residue at $w=\lambda $ combines with the second term
to produce $V'(\lambda )-W_0(\lambda )=y(\lambda )$ while we can replace $V'(w)$ by $y(w)$ when evaluating the residue at infinity
due to the asymptotic conditions. That is, we obtain
\bea
&&y(\lambda )\oint_{{\cal C}_{{\cal D}_\mu }}\frac{d\mu }{2\pi i}\frac{d{\cal G}(\lambda ,\mu )}{d\lambda }
\frac{1}{y(\mu )(\mu -\mu_\alpha)^k}+
\oint_{{\cal C}_\infty}\frac{dw}{2\pi i}\,\frac{y(w)}{\lambda -w}\oint_{{\cal C}_{{\cal D}_\mu }}\frac{d\mu }{2\pi i}
\frac{d{\cal G}(w,\mu )}{dw}\frac{1}{y(\mu )(\mu -\mu_\alpha)^k}
\nonumber
\\
&=&
\oint_{{\cal C}_{{\cal D}_w}}\frac{dw}{2\pi i}\frac{y(w)}{\lambda -w}\oint_{{\cal C}_{{\cal D}_\mu }}\frac{d\mu }{2\pi i}
\frac{d{\cal G}(w,\mu )}{dw}\frac{1}{y(\mu )(\mu -\mu_\alpha)^k},
\nonumber
\eea
where the contour ordering is such that $\lambda >{\cal C}_{{\cal D}_w}>{\cal C}_{{\cal D}_\mu }$. We now want to push the
integration contour for $\mu $ through the integration contour for $w$. After it, the obtained integral over $\mu $ vanishes
as the integrand is then analytic everywhere outside ${\cal C}_{{\cal D}_\mu }$. Thus, contributions come only from the
pole at $w=\mu $ of $d{\cal G}(w,\mu )$, which contributes when pushing the contour ${\cal C}_{{\cal D}_\mu }$ through ${\cal C}_{{\cal D}_w}$,
and from the multiple-valuedness of $d{\cal G}(w,\mu )$ w.r.t. the variable
$\mu$. Note, however, that these latter contributions are always proportional to
$d\omega_i(w )=\frac{ H_i(w)d\lambda }{\ty(w )}$.
That is, we have
\bea
&&\oint_{{\cal C}_{{\cal D}_w}}\frac{dw}{2\pi i}\frac{y(w)}{\lambda -w}\frac{1}{y(w)(w-\mu_\alpha)^k}
\nonumber
\\
&&\quad\quad+\hbox{const\,}\cdot\oint\oint_{\lambda >{\cal C}_{{\cal D}_\mu }>{\cal C}_{{\cal D}_w}}\frac{dw}{2\pi i}\frac{d\mu }{2\pi i}
\frac{y(w)}{\lambda -w} \frac{ H_i(w)}{\ty(w)}\frac{1}{y(\mu )(\mu -\mu_\alpha)^k},
\nonumber
\eea
where the point $\lambda $ lies outside the integration contour in the first term and the integral over $w$ in
the last term vanishes because the integrand
$$
\frac{y(w) H_i(w)}{(\lambda -w)\ty(w)}=\frac{M(w) H_i(w)}{\lambda -w}
$$
is obviously regular everywhere inside the contour ${\cal C}_{{\cal D}_w}$ (recall that both $\lambda $ and $\mu $ are now outside
this contour). Upon integration, the first term obviously produces $(\lambda -\mu_\alpha)^{-k}$, which completes the
first part of the proof.

Next, note that conditions (\ref{A-cycle-F}) hold automatically for any function $f(\lambda )$,
having singularities only at $\mu_\alpha$, transformed by the operator $\widehat{d{\cal G}}$,
(\ref{normG}) due to the normalization properties of the kernel $d{\cal G}(\lambda ,\mu )$ (\ref{*dE}).
Therefore, the result of the action of this operator can be always presented as the linear
combination of the functions $\frac{\d \mu_\alpha}{\d V(\lambda )}$ and
$\frac{\d M_\alpha^{(k)}}{\d V(\lambda )}$, which completes the proof of
formula (\ref{*1*}).

\subsection{Calculations in genus one
\label{ss:4.2}}

Now we invert the loop equations for genus $h=1$ and integrate them to obtain
the genus one free energy. We need, in this case, only
$\chi_\alpha^{(1)}(\lambda)$ and $\chi_\alpha^{(2)}(\lambda)$ (see
(\ref{Wstr})) which we already have, (\ref{basis2}), and
eq.(\ref{4.11}) reads
\be
(\widehat{K}-2W_0(\lambda))W_1(\lambda)  =  \dV W_0(\lambda). \label{loop1}
\ee
Given $W_0(\lambda)$ (\ref{W0}), the r.h.s. becomes
\bea
\dV W_0(\lambda)    &=& -\frac{3}{16}\sum_{\alpha=1}^{2n}\frac{1}{(\lambda-\mu_\alpha)^2}
            - \frac{1}{8}\sum_{\alpha,\beta=1 \atop \alpha<\beta}^{2n}
              \frac{1}{(\lambda-\mu_\alpha)(\lambda-\mu_\beta)}
             \nonumber\\
           &&+ \frac{1}{4}\tilde y(\lambda)\sum_{\alpha=1}^{2n}\frac{1}{\lambda-\mu_\alpha}
               M_\alpha^{(1)} \dmul \nonumber\\
             &=& \frac{1}{16}\sum_{\alpha=1}^{2n}\frac{1}{(\lambda-\mu_\alpha)^2}
          - \frac{1}{8}\sum_{\alpha,\beta=1 \atop \alpha<\beta}^{2n}
                 \frac{1}{\mu_\alpha-\mu_\beta}\left(\frac{1}{\lambda-\mu_\alpha}-
                 \frac{1}{\lambda-\mu_\beta}\right)
              \nonumber\\
           &&+ \frac{1}{4}\sum_{\alpha=1}^{2n}
           \frac{{\cal L}_{\alpha}(\mu_\alpha)}{\lambda-\mu_\alpha}.
 \label{dW0}
\eea
Here we took into account that regular parts coming from
$\frac{\lambda^l}{\lambda-\mu_\alpha}$, $l=1,\ldots,n-2$, vanish
for $W_0(\lambda,\lambda)=\dV W_0(\lambda)$ to satisfy the correct asymptotic behavior, and
we can just replace $\lambda^l$ by $\mu_\alpha^l$ in numerators of such expressions.
The result for the one-point resolvent of genus one with $n$ cuts can now be
easily obtained using Eq. (\ref{basis2}),
\bea
W_1(\lambda)  &=& \frac{1}{16}\sum_{\alpha=1}^{2n}\chi_\alpha^{(2)}(\lambda)
                  - \frac{1}{8}\sum_{1\leq\alpha<\beta\leq 2n}
           \frac{1}{\mu_\alpha-\mu_\beta}\left(
           \chi_\alpha^{(1)}(\lambda)-\chi_\beta^{(1)}(\lambda)\right) \nonumber\\
            && + \frac{1}{4}\sum_{\alpha=1}^{2n}
                 {\cal L}_{\alpha}(\mu_\alpha)\chi_\alpha^{(1)}(\lambda)
                 \nonumber\\
             &=&\ \frac{1}{16}\sum_{\alpha=1}^{2n}
             \left(-\frac{2}{3}\dV \log |M_\alpha^{(1)}| -\frac{1}{3}
          \sum_{\beta=1 \atop \beta\neq \alpha}^{2n} \dV \log |\mu_\alpha-\mu_\beta|\right)
          \nonumber
          \\
          &&- \frac{1}{8}\sum_{\alpha,\beta=1 \atop \alpha<\beta}^{2n}
                 \frac{1}{\mu_\alpha-\mu_\beta}\left(\dmul-\frac{\d\mu_\beta}{\d V(\lambda)}\right)
              \nonumber\\
           &&+ \frac{1}{4}\sum_{\alpha=1}^{2n}{\cal L}_{\alpha}(\mu_\alpha)\dmul.
\label{W1}
\eea
Now one should integrate (\ref{W1}) in order to obtain ${\cal F}_1$.
While integrating the first two terms in the r.h.s. is straightforward, the
term with the zero modes requires some more work.
Using formulas (\ref{main}) and (\ref{inverse}), we see
that the last term in (\ref{W1}) is just
$$
-\frac12\sum_{\alpha=1}^{2n}\frac{\partial}{\partial\mu_\alpha}\left(\log\det_{i,j=1,\dots,n-1}
\sigma_{j,i}\right)\dmul,
$$
i.e.,
\be
\label{F1}
{\cal F}_1=-\frac1{24}\log\left(\prod_{\alpha=1}^{2n}M(\mu_\alpha)\cdot\Delta^{4}\cdot
(\det_{i,j=1,\dots,n-1}\sigma_{j,i})^{12}\right),
\ee
where $\Delta=\prod_{1\leq\alpha<\beta\leq 2n}(\mu_\alpha-\mu_\beta)$ is the
Vandermonde determinant. This is our final answer for the genus-one partition
function\footnote{One has to compare this answer with the formula proposed
in~\cite{DW,Wit90,Wit91} for the one-loop (toric) corrections in topological
theories,
$$
\F_1(t_I)=\frac{1}{24}\log\det \left[\frac{\d^3\F_0}{\d t_\alpha\d
t_\beta\d t_\gamma} \d_X t_\gamma\right]+G(\{t_\alpha\}).
$$}.

\subsection{Genus one free energy and determinant representation
\label{ss:torus}}

Let us now discuss the expression for the genus one free
energy. First of all, notice that (\ref{F1}) reproduces the calculation of
\cite{Ak96} for the two-cut solution, up to a modular transformation
permuting $A-$ with $B-$ cycles. This should be a surprise,
since we put throughout our calculation in sect.~\ref{ss:4.2} the constraint
that $A$-periods (\ref{Sfr}) of the generating differential
(\ref{dS}) are constant under the action of the operator ${\d\over\d V(p)}$,
see (\ref{dVSi}). On the contrary, in \cite{Ak96} Akemann imposed the condition
of vanishing $B$-periods of
$dS$, corresponding to equal ``levels" $\Pi_i$ in different wells of the
potential \cite{david92} or additional minimization of the free energy
(\ref{variF}) at the saddle point w.r.t. the occupation numbers (\ref{oc}).
In fact, since neither the answer (\ref{F1}) nor intermediate calculations
contain any manifest dependence on particular values of the
periods, one can equally put all $B$-periods (during the calculation of \cite{Ak96})
fixed to be arbitrary non-vanishing constants
and, therefore, come to a modular transformed counterpart of our choice of the
normalizing cycles.

Under condition of constant $B$-periods (\ref{Dvdual})-(\ref{Dvdual1}),
as it was stressed in
\cite{KMT}, the matrix $\sigma_{i,j}$ of the $A$-periods of ${x^idx\over \ty(x)}$
is replaced by the matrix of the corresponding $B$-periods. This is
the only difference with the result of \cite{Ak96}; certainly formula (\ref{F1})
reproduces the answer of \cite{Kos,DST}\footnote{If restoring in \cite{Kos}
the determinant term $\det \sigma$, omitted from the answer.}
for generic multi-cut solution.

The fact, that the only result of interchanging $A$- and
$B$-cycles is the interchanging of the corresponding periods in $\det \sigma$
implies that $e^{{\F}_1}$ is a {\em density} and not a scalar function on moduli
space of the curves. Indeed, when exchanging $A$- and $B$-cycles,
$\det \sigma_{i,j}$ is multiplied by $\det
\tau_{ij}$ -- the determinant of the period matrix of the curve. In order to compensate
this factor, $e^{{\F}_1}$ must be transformed under such transformation
with the additional factor
$\left(\det\tau_{ij}\right)^{1/2}$, as follows from (\ref{F1}). Then one immediately
comes to the above observation: exchange of $A$- and $B$-cycles results
only in replacing the corresponding matrix $\sigma_{i,j}$ in (\ref{F1}). Note that such
behavior of $e^{{\F}_1}$ indicates that it is a section of determinant bundle
${\rm DET}\bar\d$ over the moduli space, where the $\bar\d$-operator acts on the
sections of a non-trivial bundle on a complex curve of matrix model.
One can find that determinant $\det '\bar\d_j$ of the $\bar\d$-operator (with some
fixed basis of the zero modes), acting on $j$-differentials, is
proportional to $\det\sigma$ for $j=0,1$ but for other values of $j$
it typically does not contain
the factor $\det\sigma$, still transforming non-trivially under exchange of $A$- and
$B$-cycles. It was proposed in \cite{DST} that, in order to match
the proper behavior under modular transformations, the operator $\bar\d_j$
should act on twisted bosons on hyperelliptic curves, then $e^{{\F}_1}$ actually
equals to its determinant. Besides, one also needs to add some corrections from
the star operators \cite{DST,Moore} that do not contain $\det\sigma$
factors and cannot be restored by modular covariance
of the answer; these are necessary to obtain the correct result (\ref{F1}).

There is another important point that differs between our formula and the result
of \cite{Ak96}. Namely, while in \cite{Ak96} it was possible to add any constant to
the final result, not spoiling the solution to the loop equation,
we can add to (\ref{F1}) an arbitrary
{\em function} of occupation numbers $S_i$, since in contrast to \cite{Ak96}
with no free parameters, we keep $S_i$'s arbitrary.

One can partially fix this arbitrary function in the free energy by imposing
requirement of smooth behavior of ${\cal F}_1$ under degenerations of the
surface. To this end, let us shrink one of the cuts, e.g. bring $\mu_2$ to
$\mu_1$. Setting $\mu_2-\mu_1=\epsilon\to 0$, we can easily check that
\be
{\cal F}_1^{(n)}\sim -{1\over 24}\log\left[
\epsilon^4\prod_{\alpha=3}^{2n}(\mu_1-\mu_\alpha)\prod_{i=1}^{m-n}
(\mu_1-\lambda_i)^2\right]+{\cal F}_1^{(n-1)}+O(\epsilon)
\ee
In order to compensate the first term that spoils the smooth degeneration,
one suffices to add
\be\label{addsm}
+{1\over 12}\prod_{i=1}^n \log S_i,\ \ \ \ S_n\equiv t_0-\sum_{i=1}^{n-1}S_i
\ee
Note that this term is out of control in the conformal field theory approach
of \cite{Kos,DST}. On the other hand, it could be also compared with the matrix
model calculations of \cite{David,KMT}. An arbitrary function of $S_i$ comes there
from different normalizations of the matrix integral. In particular, the
normalization in \cite{David,KMT} corresponds just to (\ref{F1}) without
adding (\ref{addsm}).

This is, in fact, a general phenomenon in the matrix model calculations:
for any genus the
only source for the singular contribution comes from degenerate geometry
of curves and is related with normalization factor in the matrix integral, that is,
the volume of (the orbit of) the unitary group. Indeed, the integral itself is a
Taylor series
in $S_i$'s (see formula (4.8) in \cite{KMT}), while the unitary group volume
\cite{versus} contributes with the factor
\be
\prod_i \left(\prod_{l=1}^{S_i/\hbar}\Gamma (l)\right)\equiv\prod_i
G_2(S_i/\hbar)
\ee
where $G_2(x)$ is the Barns function, \cite{Barns}. Now
using the asymptotic expansion for the $\Gamma$-function at large values
of argument and formulas relating the $\Gamma$-function and the Barns function
(see, e.g., \cite{WW}), one finds the asymptotic expansion of the Barns function
\cite{Barns,d}
\be
G_2(N)=\log\left(\prod_{l=1}^N\Gamma(l)\right)={S^2\over 2}\log N -{1\over 12}\log
N-{3\over 4}N^2 +{1\over 2}N\log 2\pi+\zeta'(-1)+\sum_{h=2}{B_{2h}\over
4h(h-1)}{1\over N^{2h-2}}
\ee
where $B_{2h}$ are Bernoulli coefficients and $\zeta (s)$ is the Riemann
$\zeta$-function. Thus, one
obtains that the singular contribution in genus $h$ is \cite{KMT}
\be
{\cal F}_h=\sum_i^n {B_{2h}\over
4h(h-1)}{1\over S_i^{2h-2}}\ \ \ \ h\ge 2
\ee
This is the simplest way to pick up the singular contribution, although it
can be also done, genus by genus, by direct solving the loop equations,
like it was demonstrated above for the genus one case.

Now let us turn to another important issue. In sect.~\ref{ss:WDVV}, we
obtained formula (\ref{Fphi3}), expressing the third derivatives of $\F_0$ through
the quantities $\phi_{I\alpha}$ determined in (\ref{phi}).
Since one often interprets ${\cal F}$ as the free energy of a topological
string theory, one could naturally associate the third derivative of $\F_0$ with
the tree three-point function in this theory, i.e. represent \rf{Fphi3} as
three ``propagators" $\phi_{I\alpha}$ ending at the same ``3-vertex":
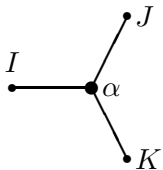
$$
\begin{picture}(90,55)(10,10)
\thicklines
\put(20,40){\line(1,0){20}}
\put(20,44){\makebox(0,0)[cb]{$I$}}
\put(51,58){\makebox(0,0)[lc]{$J$}}
\put(51,22){\makebox(0,0)[lc]{$K$}}
\put(20,40){\circle*{2}}
\put(49,58){\circle*{2}}
\put(49,22){\circle*{2}}
\put(40,40){\circle*{3}}
\put(45,38){\makebox(0,0)[cb]{$\alpha$}}
\put(40,40){\line(1,2){9}}
\put(40,40){\line(1,-2){9}}
\end{picture}
$$
In such case, one has to associate
$\F_1$ with the one-loop diagram in this topological theory, i.e. with the
propagator determinant $\det_{I,\alpha}\phi_{I\alpha}$:
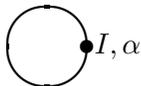
$$
\begin{picture}(90,35)(10,20)
\thicklines
\put(60,40){\oval(20,20)}
\put(70,40){\circle*{3}}
\put(72,40){\makebox(0,0)[lc]{$I,\alpha$}}
\end{picture}
$$
Calculating this determinant stems actually
to calculating the polynomial determinant
$\det_{I,\alpha} H_{I}(\mu_\alpha)$.
We already saw that, due to normalization conditions (\ref{bip})
and (\ref{ass}), the polynomials $H_{K}(\lambda)$ corresponding
to the variables $t_k$ with $k>0$ always have the coefficient $k$
at the highest term $\lambda^{n-1+k}$ while the polynomial $ H_{0}(\lambda)$
starts with unit coefficient at $\lambda^{n-1}$.
Passing from $ H_{i}(\lambda)$, corresponding to the variables $S_i$,
$i=1,\dots,n-1$, to the basis of monomials $\lambda^{i-1}$, one obtains that the
total determinant is then (up to a trivial factor $n!$)
just the total Vandermonde determinant
divided by the determinant of the transition matrix $\sigma$ (\ref{Q}). The complete
answer for the determinant of $\phi_{I\alpha}$ is then
\be
\label{xxxxx}
\det_{I,\alpha}\phi_{I\alpha}=\left(\prod\limits_{\alpha=1}^{2n}
M_{m-n}(\mu_\alpha)\right)^{-1/3}\Delta(\mu)^{-1/3}(\det\sigma)^{-1}
\ee
and the second of matching conditions (\ref{det}) stems now to the
condition of nontriviality of $\det\sigma$.\footnote{This is easy to prove.
Would be $\det\sigma=0$, one obtains
that there must exist a polynomial $P(\lambda)$ of degree less or
equal $n-2$ such that
$$
\int_{\lambda_{2i-1}}^{\lambda_{2i}}
\frac{P(\lambda)d\lambda}{\tilde y}=0\quad\hbox{for}
\quad i=1,\dots,n-1.
$$
This necessarily implies that $P(\lambda)$ has at least one zero
at each of the intervals $(\lambda_{2i-1},\lambda_{2i})$; otherwise
the combination under the integral sign is sign definite and the
integral cannot vanish. The polynomial $P(\lambda)$ must then
have at least $n-1$ zero and, having the degree not exceeding $n-2$,
must therefore vanish.}

Comparing (\ref{xxxxx}) and (\ref{F1}), one finds that powers of $\det\sigma$ and
of the Vandermonde determinant $\Delta(\mu)$ in these expressions perfectly
match, i.e. $\F_1$ is indeed proportional to the determinant, up to
non-universal pieces containing $M$ and (arbitrary function of) $S_i$.
These pieces remain due to the freedom in defining the measure in the path
integral.

Therefore, we conjecture an existence of a diagram technique for
calculating the higher genera free energy and/or generating function for the
correlators\footnote{Note that this conjectured diagram
technique is different from that of \cite{Ey}.
Note also that recent paper \cite{CEy}, where the diagram technique of \cite{Ey}
was extended from calculating resolvents to the free energy calculations,
possesses a clear disadvantage: intermediate patterns appeared are manifestly
non-symmetric w.r.t. field propagators.}.
Would such a diagram technique be constructed in full,
it opens a possibility of calculating the higher
genera/multi-point contributions in a rather effective way. Therefore, it would be
of great practical use to make further checks of the conjecture.

\subsection{Relation to topological B-model \label{ss:torus-special}}

The authors of \cite{KMT} proposed an anzatz for $\F_1$ in the two-cut
case (with absent double points). Their formula in fact comes from the
correspondence between the so called topological B-model on the local Calabi-Yau
geometry $\widehat II$ and the cubic
matrix model conjectured in \cite{DV}. However, this does not completely fix
the formula for $\F_1$, leaving room for a certain holomorphic ambiguity, which
was fixed in \cite{KMT} basically by some simplicity arguments.

First of all, introduce the quantities
$\mu_{1,2}^-\equiv \{\mu_2-\mu_1,\mu_4-\mu_3\}$,
i.e. complexified lengths of the two cuts on hyperelliptic plane,
and $\{S_1,S_2\equiv S_n=t_0-S_1\}$. Then, one expects
\be
\F_1=\frac12 \log\left(\det\left\|\frac{\d \mu_{1,2}^-}{\d S_{1,2}}\right\|\Delta(\mu)^{2/3}
(\mu_1+\mu_2-\mu_3-\mu_4)^{-1}\right).
\label{*brane}
\ee
This formula was, indeed, checked for a few first terms of expansion in
$S_i$'s \cite{KMT} and it is proven by the direct calculation in \cite{Vas}.

Below we propose a similar formula for the case of any number of cuts
(see also the details in \cite{Vas}).
Let us divide all the branching points into two ordered sets
$\{\mu^{(1)}_j\}_{j=1}^n$ and $\{\mu^{(2)}_j\}_{j=1}^n$ and
perform then a linear orthogonal transformation of
$\mu_j^{(1,2)}$ to the quantities
$\{\mu_j^+\}_{j=1}^n$ and $\{\mu_j^-\}_{j=1}^n$ by
\be
\mu_j^{\pm}=\mu^{(1)}_{j}\pm\mu^{(2)}_{j}.
\label{**1*}
\ee
Taking now $n-1$ canonical variables $S_i$, the variable $S_n=t_0-\sum_{i=1}^{n-1}S_i$,
$p$ {\em lower} times
$t_k$, $k=1,\dots,p$ \ $(0\le p\le n)$, and choosing an {\em arbitrary} set of $n+p$
branching points
$\mu_{\alpha_j}$, $j=1,\dots,n+p$, following the same logic as for (\ref{xxxxx})
(see also (\ref{diff2})), we obtain
\be
\det\left\|\frac{\d \{\mu_{\alpha_j}\}}{\d \{S_i,S_n,t_k\}}\right\|=
\frac{\Delta(\mu_{\alpha_j})\cdot (\det\sigma)^{-1}}
{\prod\limits_{j=1}^{n+p}M_{\alpha_j}^{(1)}\prod\limits_{j=1}^{n+p}
\left(\prod\limits_{\beta\ne\alpha_j}^{2n}
(\mu_{\alpha_j}-\mu_\beta)\right)}
\label{**2*}
\ee
with the {\em same} matrix $\sigma$ for any choice of the set of indices
$\{\alpha_{j}\}_{j=1}^{n+p}$ and any
number $p$ of canonical times $t_k$ (but only for $0\le p\le n$).
Set all $M_\alpha^{(1)}\equiv1$;
the Vandermonde determinant $\Delta(\mu_{\alpha_j})$ then combines with the rational
factors in the denominator to produce
$(-1)^{\sum_{j=1}^n\alpha_j}\Delta(\overline{\mu_{\alpha_j}})/\Delta(\mu)$, where
$\Delta(\overline{\mu_{\alpha_j}})$ is the Vandermonde determinant
for the supplementary set of $n-p$ branching points not entering
the set $\{\mu_{\alpha_j}\}_{j=1}^{n+p}$ whereas $\Delta(\mu)$
is the total Vandermonde determinant. In particular, when $p=0$,
splitting $\mu_\alpha$ as in (\ref{**1*}) and using
formulas (\ref{F1}) and (\ref{**2*}), we have
\be
\F_1\left|_{M_{\alpha}^{(1)}\equiv1}=\frac12
\log\left(\det\left\|\frac{\d \{\mu_j^-\}}{\d\{S_i,S_n\}}\right\|
\Delta(\mu)^{2/3}\Delta^{-1}(\mu_j^+)\right)\right.,
\label{F1brane}
\ee
where the additional Vandermonde determinant is taken w.r.t.
the supplementary variables $\mu_{j}^+$. In the two-cut case
it reproduces (\ref{*brane}).

\section*{Acknowledgments}

Our work is partly supported by Federal Program of the Russian Ministry of
Industry, Science and Technology No 40.052.1.1.1112 and by the grants:
RFBR 03-02-17373 (L.Ch. and A.Mar.), RFBR 04-01-00646 (A.Mir.), RFBR 04-02-16880 (D.V.),
Grants of Support for the Scientific
Schools 2052.2003.1 (L.Ch.), 1578.2003.2 (A.Mar), 96-15-96798 (A.Mir.),
and by the Program Mathematical Methods of
Nonlinear Dynamics (L.Ch.). The work of A.Mar. was also supported by the
Russian Science Support Foundation.

\end{document}